\documentclass[11pt,a4paper]{article}

\DeclareFontShape{OMX}{cmex}{m}{b}{<-> cmexb10}{}
\SetSymbolFont{largesymbols}{bold}{OMX}{cmex}{m}{b}

\usepackage{amsmath}
\usepackage{amssymb}
\usepackage{amsfonts}

\usepackage[utf8]{inputenc}
\DeclareUnicodeCharacter{2212}{-} 

\usepackage{graphicx}
\usepackage[autostyle]{csquotes}
\usepackage[linktocpage]{hyperref}
\usepackage{fullpage}
\usepackage{slashed}
\usepackage{mathtools}
\usepackage{indentfirst}
\usepackage{bm}
\usepackage{mathrsfs} 
\usepackage{authblk}
\usepackage{tikz}
\usepackage{float}

\usepackage{fixmath}

\allowdisplaybreaks

\newtheorem{theorem}{Theorem}[section]

\newtheorem{Defn}[theorem]{Definition}
\newtheorem{Thm}[theorem]{Theorem}
\newtheorem{Prop}[theorem]{Proposition}

\newtheorem{Rmk}[theorem]{Remark}
\newtheorem{Conj}[theorem]{Conjecture}

\newcommand{\nat}{\mathbb{N}}
\newcommand{\intg}{\mathbb{Z}}
\newcommand{\rational}{\mathbb{Q}}
\newcommand{\real}{\mathbb{R}}
\newcommand{\complex}{\mathbb{C}}

\newcommand{\rarw}{\rightarrow}

\newcommand{\lb}{\left(}
\newcommand{\rb}{\right)}
\newcommand{\lsb}{\left[}
\newcommand{\rsb}{\right]}
\newcommand{\lac}{\left\{}
\newcommand{\rac}{\right\}}

\newcommand{\ptl}{\partial}
\newcommand{\grd}{\nabla}

\newcommand{\RN}[1]{%
  \textup{\uppercase\expandafter{\romannumeral#1}}}

\newcommand{\rotsimeq}{\rotatebox[origin=c]{90}{$\simeq$}}

\title{Quantum invariants of 3-manifolds and links: a review}
\author{John Chae}
\affil{yjchae@formerstudents.ucdavis.edu}
\date{}  

\begin{document}

\maketitle

\begin{abstract}
We review the recent developments of quantum invariants of 3-manifolds and links: $\hat{Z}$ and $F_L$. They are $q$-series invariants originated from mathematical physics. They exhibit rich features, for example, quantum modularity, infinite dimensional Verma module structures and knot-quiver correspondence. Furthermore, they have connections to other topological invariants. We also provide a review of an extension of the above series invariants to Lie superalgebras.
\end{abstract}

\tableofcontents

\section{Introduction}

Topological quantum field theories (TQFTs) have been a fruitful source of interaction between physics and topology. From one to four dimensions, TQFTs have provided physical realizations of topological invariants or predicted new ones. Examples include the colored Jones polynomials and HOMFLY-PT polynomials of links~\cite{W1,NRZ}, as well as the Donaldson and Seiberg-Witten invariants of smooth four-manifolds~\cite{W3,W4}. In three dimensions, $SU(2)$ Chern-Simons TQFT predicted the Witten-Reshetikhin-Tureav (WRT)-invariant of 3-manifolds~\cite{W1}. The introduction of this invariant motivated a rigorous construction via the quantum group $U_q (sl(2,\complex))$ and its representations~\cite{RT,RT2} (see \cite{KM} for a review). This established a gateway into quantum topology from the mathematics side.
\newline

			%
			%
					%
			%
						%
			%
				%
			%
			%
			%
				%

On the mathematics side, TQFT was axiomatized in \cite{At,Se} (see \cite{F} for a review), and its breadth and depth have since been greatly enriched. Axiomatic TQFT synthesizes topology, quantum algebra, representation theory, and category theory. One direction of advancement in TQFTs has been the construction of extended TQFTs, which introduced higher categories into the scene~\cite{BD,Pries, Freed2}. There has also been progress in the classification of such TQFTs~\cite{L}. Another line of development involves the construction of non-semisimple TQFTs associated with various quantum groups. Non-semisimple invariants of manifolds first appeared through the ADO polynomials of links~\cite{ADO} and their quantum group formulation in \cite{Mu2}. In three dimensions, this type of TQFT utilizes non-semisimple categories~\cite{GKP2} and the modified quantum dimension~\cite{GPT,GKP1}.A non-semisimple TQFT has led to a new quantum invariant of links and 3-manifolds, called the CGP invariant~\cite{CGP}. One of the advantages of non-semisimple invariants is that they can distinguish manifolds that semisimple invariants cannot, and they yield nonzero results in cases where semisimple invariants vanish. Furthermore, the underlying quantum groups of these TQFTs have been generalized to quantum supergroups~\cite{GP1,GP2,H}.
\newline

Another rich source of interaction between physics and topology is the categorification program~\cite{CF} (see \cite{A,G,NO,S} for reviews). It has not only deepened the understanding of quantum invariants of manifolds, but also provided powerful new tools. In the case of link polynomials, many have been shown to be graded Euler characteristics of homology theories. For example, the Alexander polynomial, Jones polynomial, and HOMFLY-PT polynomial are the Euler characteristics of knot (or link) Floer homology~\cite{OS,OS2,JR}, Khovanov (co)homology~\cite{K1, K2} and Khovanov-Rozansky homology~\cite{KR2}, respectively. Furthermore, the quantum group itself has been categorified. This, combined with the quantum Weyl group, has led to a different approach to computing link polynomials~\cite{KL2,La}.
\newline

From the physics perspective on categorification, string theory has played a vital role. Beginning with knot polynomials~\cite{OV}, the first physical realization of knot homology was achieved in \cite{GSV}. It provided physical interpretations of Khovanov homology, Khovanov-Rozansky $sl(N)$ homology~\cite{KR1} and knot Floer homology. Furthermore, it predicted the existence of a categorification of the HOMFLY-PT polynomial—an unexpected development from the mathematics side~\cite{K4}. In the case of colored HOMFLY homology, a detailed investigation for torus links revealed structural properties and differentials \cite{GNSSS}. Additionally, an application of a spectral sequence to a 4-dimensional supersymmetric quantum field theory (QFT) was accomplished. Subsequently, a gauge-theoretic realization of Khovanov homology using a brane system in string/M-theory was introduced in \cite{W2} (see \cite{W9,W10} for reviews).
\newline

A major challenge of the categorification program has been categorifying the $sl(2,\complex)$ WRT invariant of closed 3-manifolds $Y$. The invariant is defined at roots of unity and does not exhibit a manifest integrality property that would allow it to be interpreted as the Euler characteristic of a homology theory. A strategy for its categorification was proposed in \cite{K3,EQ}. On the physics side, a 3-dimensional supersymmetric quantum field theory (QFT) arising from six dimensions predicted the existence of a power series with integral coefficients associated with the WRT invariant~\cite{GPV,GPPV}. This $q$ series, denoted by $\hat{Z}_b $, is labeled by $Spin^c$ structures $b$ on $Y$, representing a vast generalization of \cite{LZ}. . Notably, the appearance of $Spin^c$ structures is novel, as the original definition of the WRT invariant does not involve such structures. Moreover, $\hat{Z}_b$ itself is a topological invariant of $Y$, implying that there are multiple invariants associated with $Y$ distinguished by the choice of $b$. It was conjectured that the WRT invariant of $Y$ can be expressed as a linear combination of the
$\hat{Z}_b $'s. This decomposition was proven for a particular class of 3-manifolds in \cite{Mu}.
\newline

Importantly, it was also conjectured that $\hat{Z}_b $ is the graded Euler characteristic of a homology theory that would provide the desired categorification of the
WRT invariant. From the physics point of view, $\hat{Z}_b$ is the nonperturbative partition function for $SL(2,\complex)$ complex Chern-Simons theory on $Y$. 
\newline

A generalization to 3-manifolds with torus boundary—particularly plumbed knot complements—was achieved in \cite{GM}. This led to the definition of a two-variable series invariant, $F_K(x,q)$ for the complement of a knot $K$. The series $F_K(x,q)$ provided access to $\hat{Z}_b$ for closed manifolds beyond plumbed manifolds via  Dehn surgery formulas. Following the introduction of the $q$-series $\hat{Z}_b $ and $F_K$, there has been extensive development. For example, there are extensions to higher rank Lie groups~\cite{P1}, the discovery of a quantum modularity property~\cite{CCFGH, CCKPS}, connections to other (geometric) topological invariants~\cite{GPP,C2}, an R-matrix formulation and generalizations to links, denoted $F_L$~\cite{P2,P3,Gr} and a quiver formulation~\cite{EGGKPSS}.
\newline

Motivated by $\hat{Z}_b$, a variety of extensions of $\hat{Z}_b (q)$ and $F_K$ have been introduced. For example, a two variable refinement $\hat{\hat{Z}}_b (q,t)$ for negative definite plumbed 3-manifolds was defined in \cite{AJK}. This invariant originates from lattice cohomology theory and reduces to $\hat{Z}_b (q)$ when $t=-1$. The quantum modularity aspects of $\hat{\hat{Z}}_b (q,t)$ was explored in \cite{LM}. A generalization of $\hat{\hat{Z}}_b (q,t)$ to knot complements was presented in \cite{AJP}. Another extension was introduced in \cite{MT}, where a set of formal series denoted by $Y(q)$ is associated with higher-rank Lie groups and generalized $Spin^c$ structures. It was shown that, among $Y(q)$ series, the one invariant under the Weyl group action coincides with $\hat{Z}_b (q)$, thereby demonstrating the uniqueness of $\hat{Z}_b$.
\newline

An algebraic extension was introduced in \cite{FP}, namely, a $q$-series invariant associated with Lie superalgebras. In case of $sl(2|1)$, the series was denoted by $\hat{Z}_{b,c} (q)$ and carries two labels $(b,c) \in Spin^c (Y) \times Spin^c (Y)$. For a class of 3-manifolds called plumbed manifolds $Y(\Gamma)$ , it was shown that $\hat{Z}_{b,c}$ decomposes a quantum invariant of $Y(\Gamma)$ constructed in \cite{H}. From the physics viewpoint, string/M-theory predicted the existence of the topological invariant $\hat{Z}_{b,c} (q)$. Furthermore, the super $\hat{Z}_{b,c} (q)$ was generalized to plumbed knot complements in \cite{C5}, leading to a three-variable series $F_K(y,z,q)$ that exhibits distinctive features compared to $F_K(x,q)$.
\newline

In this review article, we provide an overview of the developments of the $q$-series invariants $\hat{Z}_b (q), F_K (x,q)$ and its link generalization $F_L$ by various aspects of these invariants. Moreover, we also survey their extensions to supergroups.  
\newline

\noindent\textbf{Organization of the paper.} In Section 2 we describe the series invariant $\hat{Z}_{b}$ for closed 3-manifolds, its underlying physics, properties, effects of line operator insertions and relations to other invariants.\\
\indent In Section 3 we describe the series invariant $F_L $ for complement of links, its R-matrix formulation, surgery formulas, and its connections to the ADO polynomials and the quiver theory.  \\
\indent In Section 4 we review an  the series invariant super $\hat{Z}_{b,c}$ associated with a Lie superalgebra for closed 3-manifolds and its underlying physics.\\
\indent Finally, in Section 5 we summarize a three variable series $F_K(y,z,q)$ for complements of plumbed knots and its relation to the super $\hat{Z}_{b,c}$. We list open problems for future directions.\\
\newline

\noindent\textbf{Acknowledgment} I am grateful to Sergei Gukov for valuable comments on a draft of this paper.

%
%
%

\section{Series invariant for closed 3-manifolds}

As mentioned in the introduction, a major challenge in the categorification program has been categorifying the WRT invariant of 3-manifolds. The goal is to define homology groups for closed, oriented 3-manifolds whose graded Euler characteristic equals the WRT invariant or an invariant closely related to it. This homology theory can be regarded as a 3-manifold analogue of Khovanov homology. A physics approach to the problem was introduced in \cite{GPV,GPPV}. Specifically, generalizing the result of \cite{LZ}, the existence of a $q$-series invariant for closed oriented 3-manifolds $Y$  denoted by $\hat{Z}_{b}(Y;q)$, exhibiting integrality properties was conjectured in \cite{GPV, GPPV}. For $Y$ with $b_{1}(Y)=0$ (i.e $Y=\rational HS^3$) and every $Spin^c (Y)$ structure $b$, 
$$
\Delta_{b} \in \rational,\quad c \in \intg_{+},\quad \hat{Z}_{b}(Y;q) \in \frac{1}{2^c}q^{\Delta_{b}} \intg[[q]], \quad |q|<1.
$$
It is a convergent $q$-series in the interior of the complex unit disc. It is conjectured that the WRT invariant of $Y$ decomposes in terms of $\hat{Z}_b (q)$:
\begin{Conj}({\cite{GPPV}})\, Let $Y$ be a closed 3-manifold with $b_{1}(Y) = 0$. Let $Spin^c (Y)$ be the set of $Spin^c$ structures on $Y$, with the action of $\intg/2$ by conjugation. Set
$$
T : = Spin^c (Y)/ \intg_2.
$$
The radial limit $\lim_{q \rarw e^{i2\pi/k}} \hat{Z}_{b}(q)$ exists and in this limit, the WRT invariant of $Y$ decomposes as a linear combination of $\hat{Z}_{b}(q)$:
\begin{equation*}
WRT[Y;k]= \frac{1}{i\sqrt{2k}} \sum_{a,b \in T} e^{i2\pi k\, lk(a,a)}\frac{1}{|W_{b}|} S_{ab} \, \hat{Z}_{b}(q) \Big\vert_{q \rarw e^{\frac{i2\pi}{k}}}
\end{equation*}
$$
S_{ab}= \frac{e^{i2\pi k\, lk(a,b)}+e^{-i2\pi k\, lk(a,b)}}{|W_{b}|\sqrt{| H_1(Y;\intg) |}}, \qquad lk : Tor\, H_{1}(Y;\intg) \times Tor\, H_{1}(Y;\intg) \rarw \rational/\intg,
$$
where $W_{x}=Stab_{\intg_2}(x)$ is $\intg_2$ if $x=-x$ and is 1 otherwise; $lk$ is the linking pairing.  
\end{Conj}
Furthermore, $\hat{Z}(Y;q)$ is supposed to admit a categorification
$$
\chi[\mathcal{H}^{i,j}_{BPS}(Y;b)]=\hat{Z}_{b}[Y;q]= \sum_{i,j} (-1)^i\, q^j\, \text{dim}\, \mathcal{H}^{i,j}_{BPS}(Y;b)
$$
This homology groups $\mathcal{H}^{i,j}_{BPS}(Y;b)$ are claimed to be the desired homology groups categorifying the WRT invariant.

\subsection{Plumbed manifolds}

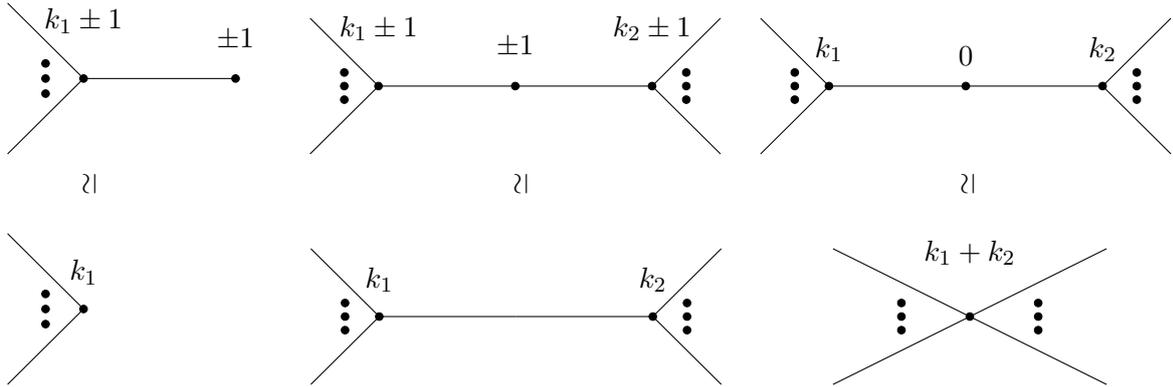
\begin{figure}[t!]
\begin{tikzpicture}[scale=1]
\tikzstyle{every node}=[draw,shape=circle]

\draw (0,0) node[circle,fill,inner sep=1pt,label=above:$k_1 \pm 1$](){};
\draw (0,0)  -- (2,0) node[circle,fill,inner sep=1pt,label=above:$\pm 1$](){};

\draw (0,0)  -- (-1,1); 
\draw (0,0)  -- (-1,-1); 

\draw (-0.5,0.2) node[circle,fill,inner sep=1pt](){};
\draw (-0.5,0) node[circle,fill,inner sep=1pt](){};
\draw (-0.5,-0.2) node[circle,fill,inner sep=1pt](){};

\end{tikzpicture}
\quad
\begin{tikzpicture}[scale=0.9]
\tikzstyle{every node}=[draw,shape=circle]

\draw (0,0) node[circle,fill,inner sep=1pt,label=above:$ \pm 1$](){};
\draw (0,0)  -- (-2,0) node[circle,fill,inner sep=1pt,label=above:$k_1 \pm 1$](){};
\draw (0,0)  -- (2,0) node[circle,fill,inner sep=1pt,label=above:$k_2 \pm 1$](){};

\draw (2,0)  -- (3,1); 
\draw (2,0)  -- (3,-1);

\draw (-2,0)  -- (-3,1); 
\draw (-2,0)  -- (-3,-1);

\draw (-2.5,0.2) node[circle,fill,inner sep=1pt](){};
\draw (-2.5,0) node[circle,fill,inner sep=1pt](){};
\draw (-2.5,-0.2) node[circle,fill,inner sep=1pt](){};

\draw (2.5,0.2) node[circle,fill,inner sep=1pt](){};
\draw (2.5,0) node[circle,fill,inner sep=1pt](){};
\draw (2.5,-0.2) node[circle,fill,inner sep=1pt](){};

\end{tikzpicture}
\quad
\begin{tikzpicture}[scale=0.9]
\tikzstyle{every node}=[draw,shape=circle]

\draw (0,0) node[circle,fill,inner sep=1pt,label=above:$ 0$](){};
\draw (0,0)  -- (-2,0) node[circle,fill,inner sep=1pt,label=above:$k_1$](){};
\draw (0,0)  -- (2,0) node[circle,fill,inner sep=1pt,label=above:$k_2 $](){};

\draw (2,0)  -- (3,1); 
\draw (2,0)  -- (3,-1);

\draw (-2,0)  -- (-3,1); 
\draw (-2,0)  -- (-3,-1);

\draw (-2.5,0.2) node[circle,fill,inner sep=1pt](){};
\draw (-2.5,0) node[circle,fill,inner sep=1pt](){};
\draw (-2.5,-0.2) node[circle,fill,inner sep=1pt](){};

\draw (2.5,0.2) node[circle,fill,inner sep=1pt](){};
\draw (2.5,0) node[circle,fill,inner sep=1pt](){};
\draw (2.5,-0.2) node[circle,fill,inner sep=1pt](){};

\end{tikzpicture}

\begin{tikzpicture}[scale=1]
\hspace{1cm}
\rotsimeq

\hspace{5.5cm}

\rotsimeq

\hspace{5.7cm}

\rotsimeq
\end{tikzpicture}

\begin{tikzpicture}[scale=1]
\tikzstyle{every node}=[draw,shape=circle]

\draw (0,0) node[circle,fill,inner sep=1pt,label=above:$k_1$](){};

\draw (0,0)  -- (-1,1); 
\draw (0,0)  -- (-1,-1); 

\draw (-0.5,0.2) node[circle,fill,inner sep=1pt](){};
\draw (-0.5,0) node[circle,fill,inner sep=1pt](){};
\draw (-0.5,-0.2) node[circle,fill,inner sep=1pt](){};

\end{tikzpicture}
\hspace{2.3cm}
\begin{tikzpicture}[scale=0.9]
\tikzstyle{every node}=[draw,shape=circle]

\draw (0,0)  -- (-2,0) node[circle,fill,inner sep=1pt,label=above:$k_1$](){};
\draw (0,0)  -- (2,0) node[circle,fill,inner sep=1pt,label=above:$k_2 $](){};

\draw (2,0)  -- (3,1); 
\draw (2,0)  -- (3,-1);

\draw (-2,0)  -- (-3,1); 
\draw (-2,0)  -- (-3,-1);

\draw (-2.5,0.2) node[circle,fill,inner sep=1pt](){};
\draw (-2.5,0) node[circle,fill,inner sep=1pt](){};
\draw (-2.5,-0.2) node[circle,fill,inner sep=1pt](){};

\draw (2.5,0.2) node[circle,fill,inner sep=1pt](){};
\draw (2.5,0) node[circle,fill,inner sep=1pt](){};
\draw (2.5,-0.2) node[circle,fill,inner sep=1pt](){};

\end{tikzpicture}
\hspace{1.2cm}
\begin{tikzpicture}[scale=0.9]
\tikzstyle{every node}=[draw,shape=circle]

\draw (0,0) node[circle,fill,inner sep=1pt,label=above:$ k_1 + k_2 $](){};

\draw (0,0)  -- (2,1);
\draw (0,0)  -- (2,-1);

\draw (0,0)  -- (-2,1);
\draw (0,0)  -- (-2,-1);

\draw (1,0.2) node[circle,fill,inner sep=1pt](){};
\draw (1,0) node[circle,fill,inner sep=1pt](){};
\draw (1,-0.2) node[circle,fill,inner sep=1pt](){};

\draw (-1,0.2) node[circle,fill,inner sep=1pt](){};
\draw (-1,0) node[circle,fill,inner sep=1pt](){};
\draw (-1,-0.2) node[circle,fill,inner sep=1pt](){};

\end{tikzpicture}

\caption{Kirby-Neumann moves on plumbing trees. Move 1: blow up/down (left), move 2: absorption/desorption  (middle), move 3: fusion/fission (right).}
\end{figure}

We first review a class of 3-manifolds called plumbed manifolds. A closed oriented plumbed three-manifold $Y$ is described by a weighted graph $\Gamma$. It consists of vertices $\lac v_i \rac$ and edges. The former carry integer weights $\lac k_i \rac$ whereas the latter carry weight $1$. This plumbing graph data is summarized by an adjacency matrix $B$, which is a symmetric and its size is set by the number of vertices $s$ of $\Gamma$:
$$
B_{i,j} = \begin{cases}
k_i , \quad v_i = v_j \\
1 , \quad v_i , v_j \quad \text{connected}\\
0 , \quad \text{otherwise}
\end{cases}
$$
We assume that plumbing graphs are tree. An interpretation of $\Gamma$ is that each vertex $v_i$ represents a $S^1$-bundle over $S^2$ whose Euler number is $k_i$. The edge between two vertices represents gluing two $S^1$-bundles by cutting out a $D^2$ from each base space and attaching two $T^2$'s. Another useful interpretation is a surgery link $L(\Gamma)$ obtained by replacing a vertex by a $k_i$-framed unknot and an edge by a Hopf link between two unknots. Hence $L(\Gamma)$ is always a tree link. Applying Dehn surgery (see Section 3.5 for a review) on $L(\Gamma)$ results in the same $Y$. The first homology of $Y(\Gamma)$ is
\begin{equation}
H_1 (Y(\Gamma)) \cong \intg^s / B \intg^s .
\end{equation}  
In case $B$ is nondegenerate, $Y$ is a rational homology sphere. When $B$ is negative definite, we call $Y$ as a negative definite plumbed 3-manifold.
\newline

\indent A plumbed 3-manifold can be presented by different plumbing graphs that are related by a set of Kirby-Neumann moves in Figure 1. In \cite{Ki,N,FR}, it was shown that two plumbing graphs $\Gamma$ and $\Gamma^{\prime}$ represent the same 3-manifolds $Y(\Gamma) \simeq Y(\Gamma^{\prime})$ if and only if they are related by a sequence of the moves.   
\newline

\indent A well known class of plumbed 3-manifold is Seifert fibered manifolds. Its graph is star shaped; it consists of one central vertex of degree $\geq 2$~\footnote{Degree of a vertex is number of legs emanating from it. Degree two case is a Lens space (a special Seifert fibered manifold).} and finite number of legs attached to the central vertex. Degree of vertices on the legs are one or two. These legs are singular fibers of the manifold. The graph data can be summarized in the following way.
$$
M \lb b \bigg| \frac{a_1}{b_1} , ..., \frac{a_n}{b_n} \rb,\qquad gcd(a_i,b_i)=1
$$
$$
e = b + \sum_{i=1}^{n} \frac{a_i}{b_i} \in \rational, \qquad ( b \in \intg )
$$
where $e$ is the Euler number, $b$ is the weight of the central vertex, $n$ is the number of singular fibers and $(a_i , b_i )$ are called Seifert invariants. Their continued fraction expansions yield the weights of the vertices on the legs.
$$
\frac{b_i}{a_i} = 
k^{i}_1 -\cfrac{1}{k^{i}_2 -\cfrac{1}{\ddots - \frac{1}{k^{i}_s}}}.
$$
where $s$ depends on the singluar fibers. A vertex attached to the central vertex has weight $-k^{i}_1$ and the last vertex on the same leg has weight $-k^{i}_s$.  
\newline

\indent For negative definite plumbed 3-manifolds, $b<0$ and $0< a_i < b_i $. It was shown in \cite{NR} that sign of $e$ determines the positive or negative definiteness of the manifolds (converse also holds).
\newline
In case (1) is trivial $H_1 =0$, $Y(\Gamma)$ is an integral homology three-sphere $\intg HS^3$. In terms of Seifert data, the $\intg HS^3$ condition is
$$
e \prod_{i=1}^{n} b_i = \pm 1.
$$
This subclass of manifolds are denoted by $\Sigma ( b_1, \cdots, b_n)$.
\newline

\noindent \underline{$Spin^c$ structures} We next describe $Spin^c$ structures on plumbed manifolds $Y_{\Gamma}$~\cite{GM}. A $Spin^c$ structure on an oriented 3-dimensional manifold $Y$ is a lift of the structure group $SO(3)$ of its tangent bundle $TY$ to
$$
Spin^c (3) = Spin(3) \times_{\intg/2} S^1 \cong U(2).
$$
They always exist for low dimensional manifolds (dimension $\leq 4$). And they form an affine space over $H^2 (Y,\intg)\cong H_1 (Y)$; in other words, difference between two $Spin^c$ structures corresponds to a 2-cocycle. To characterization of $Spin^c$ structures of $Y_{\Gamma}$ in terms of $\Gamma$, we first look at a closed 4-manifold $X$ bounded by $Y_{\Gamma}$ ($\ptl X = Y_{\Gamma}$). 

$Spin^c$ structures on $X$ can be canonically identified with characteristic vectors $K \in H^2(X)$:
$$
K(v) := \left\langle v,v \right\rangle, \quad \text{for all}\quad v\in H_2 (X).
$$
And  $H_2 (X) = H^2 (X) = \intg^s$, where the Poincare duality was used and $s$ is the number of vertices of $\Gamma$ and Vert is a set of its vertices. We have
$$
K \equiv \vec{m} \qquad mod\quad 2\intg^s.
$$
where $\vec{m}$ is a vector whose components are weights $m_v$ for $v\in$ Vert. From above, we have a natural identification
\begin{equation}
Spin^c (X) \cong \vec{m} + 2\intg^s
\end{equation}
\newline
In order to pass to $Y_{\Gamma}$, we analyze the long exact sequence of cohomology groups for $Y_{\Gamma}$ and $X$,
$$
H^2 (X,\ptl X) \rightarrow H^2 (X) \rightarrow H^2 (Y_{\Gamma})  \rightarrow 0
$$
entails the $H^2(Y_{\Gamma}) \cong H_1 (Y_{\Gamma}) = \intg^s /B\intg^s = Coker(B)$. Moreover,
$$
Spin^c (X) \twoheadrightarrow Spin^c (Y_{\Gamma}).
$$
From (2), we get a canonical identification~\footnote{A proof that the identification is natural can be found in Section 4.2 in \cite{GM}.},
$$
Spin^c (Y_{\Gamma}) \cong \lb 2\intg^s + \vec{m} \rb / 2B\intg^s.
$$
This is in turn identified with
$$
Spin^c (Y_{\Gamma}) \cong \lb 2\intg^s + \vec{\delta} \rb / 2B\intg^s,\qquad \vec{\delta} = (\vec{\delta}_v )_{v\in Vert},
$$
where $\vec{\delta}_v= deg(v)$ is the degree of $v \in$ Vert.

\subsection{The series invariant}

Let $Y_{\Gamma}$ be a (weakly) negative definite plumbed manifolds with $b_1 (Y_{\Gamma}) =0$ (i.e $\rational HS^3$) and $B$ be its adjacency matrix, equivalently a linking matrix of $L(\Gamma)$~\footnote{The definition of weakly in the parenthesis is in Section 4.3 of \cite{GM}; we primarily deal with negative definite plumbed manifolds in this review.}. The $SU(2)$ colored Jones polynomial of $L(\Gamma)$ is given by~\cite{GPPV}
\begin{align*}
\begin{split}
J[L(\Gamma);\underline{q}]_{n_1,\cdots,n_L} & =\frac{2i}{\underline{q}^{1/2}- \underline{q}^{-1/2}} \prod_{v=1}^L \underline{q}^{\frac{a_v (n_{v}^2 -1)}{4}} \lb \frac{2i}{\underline{q}^{n_v/2}- \underline{q}^{-n_v/2}}\rb^{\text{deg}-1}\\
& \times \prod_{(v_1,v_2)\in E} \frac{\underline{q}^{n_1 n_2 /2}- \underline{q}^{-n_1 n_2/2}}{2i},
\end{split}
\end{align*}
where $E$ denotes a set of edges of $\Gamma$ and $\underline{q}=e^{i2\pi/k},\, k\in \intg_+$. The $SU(2)$ WRT invariant of $Y_{\Gamma}$ is
\begin{equation}
\tau (Y_{\Gamma}) = \frac{F[L(\Gamma)]}{F[L(+1)]^{b_{+}}F[L(-1)]^{b_{-}}},
\end{equation}
$$
F [L(\Gamma)]  = \sum_{n\in \lac 1,\cdots , k-1 \rac^L }  J[L(\Gamma);\underline{q}]_{n_1, \cdots, n_L} \prod_{v=1}^L \frac{\underline{q}^{n_v /2}- \underline{q}^{-n_v /2}}{\underline{q}^{1/2} - \underline{q}^{-1/2}}
$$
where $L(\pm 1)$ are one vertex plumbing graphs whose framings are $\pm 1$, $b_{\pm}$ are the number of positive/negative eigenvalues of $B$. 
\begin{Rmk} The invariant (3) is its Dehn surgery formulation.
\end{Rmk}

From (3), the $q$-series $\hat{Z}_b$ of $Y_{\Gamma}$ can be obtained as follows~\cite{GPPV}~\footnote{See Appendix A of \cite{GPPV} for the derivation.}. 
\begin{equation}
\hat{Z}_{b}[Y_{\Gamma} ;q] = (-1)^{\pi} q^{\frac{3\sigma - \sum_{v}m_v}{4}}\, \prod_{v \in Vert} PV \oint_{|z_v|=1}  \frac{d z_v}{i2\pi z_v} \lb z_v - \frac{1}{z_v} \rb^{2-\text{deg}(v)} \Theta^{-Y}_{b}(\vec{z},q),
\end{equation}
where
$$
\Theta^{-Y}_{b} = \sum_{\vec{l} \in 2B\intg^{L} + \vec{b}} q^{-\frac{(\vec{l},B^{-1}\vec{l})}{4}} \prod_{v \in Vert} z_{v}^{l_v},\quad b \in Spin^c(Y)\cong H_1(Y)
$$
$$
\pi = -\text{$\sharp$ (negative eigenvalues)},\qquad \sigma = \text{signature}(B),
$$
$$
PV= \lim_{\epsilon \rarw 0 } \frac{1}{2} \lb \oint_{|z_v|=1 + \epsilon} + \oint_{|z_v|=1 - \epsilon} \rb
$$
where $L=s$ is the number of components of $L(\Gamma)$ and $PV$ is the principal value prescription for the complex contour integral. From the perspective of the surgery link $L(\Gamma)$, (4) can be viewed as a surgery formula on a tree link~\footnote{A tree link is a link consisting of unknots and their linkings are all Hopf links.}.
\begin{Rmk} Negative definite refers to $B$ being negative definite (i.e. all its eigenvalues are negative).
\end{Rmk}
\begin{Prop} (\cite{GM}) The $\hat{Z}_b$ (4)  is invariant under the Kirby-Neumann moves in Figure 1.
\end{Prop}
\begin{Thm}(\cite{Mu}) Conjecture 1.1 holds for negative definite plumbed 3-manifolds.
\end{Thm}
\begin{Rmk} For plumbed manifolds having $b_{1}(Y) > 0$ (i.e- a graph containing a loop), (4) needs a modification~\cite{CGPS}.
\end{Rmk}
\begin{Rmk} A closely related graphs to plumbing graphs are splice diagrams. They can be converted into each other. The splice diagrams are	 useful for revealing the connection of $\hat{Z}_b$ to algebraic geometry. The series $\hat{Z}_b$ for splice diagrams was analyzed in \cite{GKS}.
\end{Rmk} 

\noindent We next describe the physics underlying (4).
\newline

\noindent \textbf{Physics Story} The physical prediction for $\mathcal{H}^{i,j}_{BPS}(Y;b)$ originates from a brane system in M-theory given by the following setup.\\
\begin{center}
\begin{tabular}{c c c c c c c}
11D Spacetime      & $\real$       &  $\times$    &  $T^{*}Y$    & $\times$  & Taub-NUT\\ 
N M5               & $\real$       & $\times$     & $Y$          & $\times$  & $D^2$ \\  
Symmetries         & $\phantom{1}$   & $\phantom{1}$  & $``U(1)_N"$  & $\times$  & $U(1)_R \times U(1)_q $         
\end{tabular}
\end{center}
where $Y$ is a compact Riemannian manifold and $``U(1)_N"$ exists if $Y$ is a Seifert fibered manifold. The appearance of $T^{*}Y=$ Calabi-Yau 3-fold is required by supersymmetry preservation for any choice of metric on $Y$ due to McLean's theorem. Furthermore, Taub-NUT space is necessary to preserve supersymmetry along $D^2$ world volume directions and the rotational symmetries $U(1)_R \times U(1)_q$. The $D^2$ is in the shape of a cigar whose circle part is the M-theory circle. The world-volume theory on the stack of M5 branes is $6d\hspace{0.1cm} \mathcal{N}= (2,0)$ theory. Dimensional reduction on $Y$ give rises to $3d \hspace{0.1cm} \mathcal{N}=2\, U(N)$ SCFT on $\real \times D^2$ denoted as $T[Y; G=U(N)]$. The symmetries $U(1)_R \times U(1)_q$ gives rise to the homological and quantum gradings on the BPS Hilbert space of $T[Y; G=U(N)]$, respectively. The boundary conditions $b$ on $\ptl D^2 = S^1$ provides the torsion grading. Therefore, we arrive at the existence of the triply-graded $\intg \times \intg \times Tor\, H_{1}(Y)/\intg_2$ homology group theory:
$$
\mathcal{H}_{BPS}(Y)\quad \cong \bigoplus_{\substack{b\, \in\, Tor\, H_{1}(Y)/\intg_2 \\ i\, \in\, \intg + \Delta_b  \\ j\, \in\, \intg}} \mathcal{H}^{i,j}_{BPS}(Y;b)
$$
The homological grading is denoted by $j$ and the shift factor $\Delta_b (Y) \in \rational$ in the quantum grading $i$ is related to the d-invariant (the correction term) of the Heegaard Floer homology $HF^{\pm}(Y)$~\cite{GPP}. In the case of $Y$ is a Seifert manifold, there is an additional grading. 
\newline	
	
Gluing two copies of the solid torus $S^1 \times D^2$ along their common boundary $S^1$, we can create a $S^1 \times S^2$. An important quantity that represents $3d\hspace{0.1cm} \mathcal{N}=2\hspace{0.1cm} U(N)$ theory on $S^1 \times S^2$ is the superconformal index $I_{sc}$ of $T[Y;U(N)]$ equivalently, its supersymmetric partition function~\cite{GPPV}: 
\begin{equation}
I_{sc}(q) = Tr_{\mathcal{H}^{BPS}_{S^2}} (-1)^F q^{R/2 + J_3} = Z_{T[Y]}(S^1 \times_q S^2),
\end{equation}
\begin{center}
$\mathcal{H}^{BPS}_{S^2}$: the BPS sector of the Hilbert space, equivalently Q-cohomology of all physical operators
\end{center}
\begin{center}
F: the fermion number\hspace{1cm} R: the generator of $U(1)_R$ symmetry
\end{center}
\begin{center}
$J_3$: the Cartan generator of the $SO(3)$ isometry of $S^2$
\end{center}
\noindent Furthermore, we let
$$
\hat{Z}_{b}(q) : = Z_{T[Y]}(S^1 \times_q D^2;\,b),
$$
where $b$ is $\mathcal{N}=(0,2)$  supersymmetric boundary condition on $T^2$; the subscript $q$ means that, as one traverses $S^1$, $D^2$ rotates around its symmetry axis by $Arg(q)$. It was conjectured that $\hat{Z}_{b}(Y;q)$ and its orientation reversed version $\hat{Z}_{b}(-Y;1/q)$ can be combined to form $I_{sc}[Y;q]$:
\begin{Conj} ({\cite{GPPV}})\, The superconformal index $I_{sc}$ of $T[Y]$ admits the following factorization.
$$ 
I_{sc}[Y;q] 	= \sum_{b\, \in\, Tor\, H_{1}(Y;\intg) / \intg_2 } |W_b|\, \hat{Z}_{b}(Y;q)\, \hat{Z}_{b}(-Y;1/q)\quad \in \intg[[q]],
$$
where $\hat{Z}_{b}(1/q)$ is an analytic continuation of $\hat{Z}_{b}(q)$ outside of a complex unit disc $|q|>1$. 
\end{Conj} 
This conjecture has a generalization through introducing an additional parameter $t$, hence $I_{sc}[Y;q,t] \in \intg[t][[q]]$, which is called the \textit{topologically twisted index} (the above conjecture can be recovered by setting $t=q^{\beta},\, \beta \in \intg$). This generalized conjecture was verified for $Y=S^3, L(p,1),\, O(-p)\rarw \Sigma_g$ in \cite{GPPV}.
\newline
%
%

\subsection{Quantum modularity}

An important feature of $\hat{Z} (q)$ is quantum modularity property. This is not manifest from (4). In \cite{CCFGH}, it was shown that $\hat{Z}_b (q)$ can be expressed in terms of quantum modular forms for Seifert fibered manifolds. This connection is realized by a certain representation of a covering group of the modular group $SL(2,\intg)$, which is called the metaplectic group $\widetilde{SL(2,\intg)}$~\footnote{It is an universal double covering group of $SL(2,\intg)$. It consists of elements of a pair $(\gamma,v)$,where $\gamma \in SL(2,\intg),\quad v:H \rightarrow \complex$ is a holomorphic function satisfying a condition. The group multiplication is $(\gamma,v)\ast (\gamma^{\prime},v^{\prime}) : = (\gamma\gamma^{\prime}, (v \circ \gamma^{\prime}) v^{\prime})$.}.  We begin with a review of a (sub)representation of $\widetilde{SL(2,\intg)}$.
\newline

Relevant subrepresentations for $\hat{Z}_b$ are the Weil representations. They are subrepresentations of the $2m$-dimensional representation $\Theta_m$ spanned by the vector $\theta_m,\, m\in \intg_{+}$, whose components are $\theta_{m,r},\, r\in \intg/2m\intg$.
\begin{equation}
\theta_{m,r}(\tau, z):= \sum_{l\in r\, mod\, 2m} q^{\frac{l^2}{4m}}e^{i2\pi zl},\qquad q=e^{i2\pi\tau}
\end{equation}
From (6), weight $3/2$ unary theta functions in the upper half plane $H$ can be defined by
\begin{equation}
\theta_{m,r}^1 (\tau) : = \frac{1}{i2\pi} \frac{\ptl}{\ptl z} \theta_{m,r}(\tau, z) \bigg|_{z=0}
\end{equation}
We next consider the group of exact divisors of $m$ denoted by $\text{Ex}_m$~\footnote{A divisor $n$ of $m$ is exact if $(n,m/n)=1$}. Its group operation is $n\ast n^{\prime}= nn^{\prime}/(n,n^{\prime})^2$. For a subgroup $K\subset \text{Ex}_m$, a subrepresentation $\Theta^{m+K}$ of $\Theta_m$ can be defined as follows. Consider a matrix 
\begin{equation}
\Omega_m (n)_{r,r^{\prime}} = \delta_{r+r^{\prime}\,(mod\, 2n)}\delta_{r-r^{\prime}\,(mod\, 2m/n)}.
\end{equation}
Using (8), projection operators can be defined
$$
P^{\pm}_{m}(n) := (1_m \pm \Omega_{m}(n))/2, \quad n \in\text{Ex}_m
$$
For subgroups $K$ not containing $m$ (non-Fricke case)~\footnote{For Fricke case, see Section 8 in \cite{CCFGH}.}, and $m$ is not divisible a square number,  we can define additional projection operators
\begin{equation}
P^{m+K} : = \lb \prod_{n\in K} P^{+}_m (n) \rb P^{-}_{m}(m),
\end{equation}
where $m+K$ denotes the pair $(m,K)$ for $K=\lac 1,n,n^{\prime},\cdots \rac$. Using (9), we define
$$
\theta^{m+K}_r = 2^{|K|} \sum_{l\in \intg/2m\intg} P^{m+K}_{r,l} \theta_{m,l}
$$
From above, we define a set $\sigma^{m+K}$ consisting of unequal (up to a sign) vectors $\theta^{m+K}_r$. This set provides a basis $\lac \theta^{m+K}_r | r\in  \sigma^{m+K}\rac$ for $\Theta^{m+K}$. We have a few remarks in order.
\begin{Rmk} In case $m$ is square free and \text{Ex}$_m = K \cup (m \ast K)$, then $\Theta^{m+K}$ is irreducible. 
\end{Rmk}
\begin{Rmk} In case $m$ is not square free ($m= p^2 r$ for some prime $p$ and square free $r$), (9) is modified (see Section 3.3 for details \cite{CCFGH}).
\end{Rmk}
\begin{Rmk} For Seifert fibered manifolds with three singular fibers, Fricke case is relevant.
\end{Rmk}
We next introduce the false theta functions and describe their relevance to $\hat{Z}$.
\newline

Let $g(z)= \sum_{n>0}a_g (n) q^n $ be a cusp form of half-integral or half-integral  weight $w$ . Its Eichler integral is defined by
\begin{align}
\begin{split}
\tilde{g}(z) &  : = \sum_{n>0} n^{1-w} a_g (n) q^n\\
             &  : = \frac{(i2\pi)^{w-1}}{\Gamma (w-1)} \int_{\tau}^{i\infty} g (\tau^{\prime}) (\tau^{\prime} - \tau )^{w-2}d\tau^{\prime}
\end{split}
\end{align}
Applying (10) to (7) leads to the false theta function, which is Eichler integral (10) of $w=3/2$ vector modular form:
\begin{align}
\begin{split}
\Psi_{m,r}(\tau) := \tilde{\theta}_{m,r}^1 (\tau) & = 2\sum_{n>0} (P^{-}_m (m))_{r,n} q^{n^2/4m} \in q^{\frac{r^2}{4m}}\intg [[q]]\\
                                                  & = \sum_{\substack{ l \in \intg  \\ l=r\, mod\, 2m}} sgn(l) q^{l^2/4m}
\end{split}
\end{align}
A crucial observation in \cite{CCFGH} was $\hat{Z}_b$ for Seifert fibered manifolds with three singular fibers $Y_{\Gamma} =M(b| \lac a_i/b_i \rac_{i=1}^3 )$ can be expressed as a linear combination of (11). The appropriate linear combination is given by
\begin{equation}
\hat{Z}_b (Y_{\Gamma}; q) = c \lb q^{\delta} \Psi^{m+K}_r + d \rb, \quad c \in \complex, \quad \delta \in\rational,  \quad d\in \intg[q],
\end{equation}
where
$$
\Psi^{m+K}_{r} : = \tilde{\theta}^{m+K,1}_r = 2^{|K|} \sum_{n\geq 0} P^{m+K}_{r,n} q^{n^2/4m}.  
$$
and $b \in Spin^c (Y)/\intg_2$. Furthermore, other data of $Y$ are
\begin{align*}
4m & = l.c.m \lb 4 \lac b_i \rac_{i=1}^n  \cup \lac \text{Denominators of}\quad CS(b)\rac_{0\neq b \in Spin^c (Y)/\intg_2} \rb\\
CS(b) & = - (b,B^{-1}b).
\end{align*}
When $Y_{\Gamma}$ is the Brieskorn sphere $\Sigma (b_1,b_2,b_3)$, where positive integers $b_1 < b_2 < b_3$ are pairwise relatively prime, there is one $\hat{Z}_b$~\footnote{Brieskorn spheres are $\intg HS^3$, hence $H_1 (\Sigma (b_1,b_2,b_3))=0$.}. The modular data and the Weil representation $m+K$~\footnote{We use $\Theta^{m+K}$ and $m+K$ notations interchangeably.} are fixed by the above parameters:
$$
m=  b_1 b_2 b_3,\qquad r=m- b_1 b_2 - b_2 b_3 - b_3 b_1, \qquad K=\lac 1,b_1 b_2,b_2 b_3, b_3 b_1 \rac.
$$
\begin{equation}
d = |\sigma^{m+K}| = \frac{1}{4} (p_1 -1)(p_2 -1)(p_3 -1),
\end{equation} 
where $d$ is the dimension of Weil representation. From the viewpoint of $SU(2)$ Chern-Simons theory on $\Sigma (b_1,b_2,b_3)$, it is the number of flat connections.
\begin{Rmk} A proof of (13) can be found in \cite{GM} (cf. Proposition 4.8).
\end{Rmk}
The above false theta function (11) is an example of quantum modular form defined in \cite{Z}. Specifically, (11) is a quantum modular form of weight $1/2$~\footnote{There is a weight change for quantum modular forms $w \rightarrow 2-w$ (see Section 7.3 in \cite{CCFGH} for details).} Quantum modular form is defined through a particular a difference between quantum modular form and its $SL(2,\intg)$ transform.
\begin{Defn} (\cite{Z}) A \textit{quantum modular form} of weight $k$ and multiplier $\chi$ on $SL(2,\intg)$ is a function $Q$ on $\rational$ such that for every $\gamma \in SL(2,\intg)$, the function $p_{\gamma} : \rational \backslash \lac \gamma^{-1}\infty \rac \rightarrow \complex$, defined by
\begin{align}
p_{\gamma} (x) & := Q(x) - Q|_{k,\chi} \gamma (x)\\
Q|_{k,\chi} \gamma & = Q \lb \frac{a\tau +b}{c\tau +d} \rb \chi(\gamma) \lb c\tau + d \rb^{-k}, \qquad \gamma = \begin{pmatrix}
a & b  \\
c & d 
\end{pmatrix}
\end{align}
has some property of continuity or analyticity for every $\gamma \in SL(2,\intg)$.
\end{Defn}
Another example of quantum modular form is Mock theta function. It plays an important role for $\hat{Z}$ of orientation reversed 3-manifolds $-Y$. We will discuss it in Section 2.7. Next, we move onto examples.
\newline

\noindent\textbf{Examples}~\footnote{See \cite{CCFGH} for additional examples.}
\newline

\noindent $Y=M \lb -1 | \frac{1}{2},  \frac{1}{3},  \frac{1}{9} \rb$. Its plumbing graph is depicted above.
\begin{figure}[h!]
\begin{center}
\begin{tikzpicture}[scale=1]
\centering
\tikzstyle{every node}=[draw,shape=circle]

\draw (0,0) node[circle,fill,inner sep=1pt,label=below:$-1$](){};
\draw (0,0)  -- (1,0) node[circle,fill,inner sep=1pt,label=right:$-9$](){};
\draw (0,0)  -- (-1,0) node[circle,fill,inner sep=1pt,label=left:$-2$](){};
\draw (0,0)  -- (0,1) node[circle,fill,inner sep=1pt,label=right:$-3$](){};

\end{tikzpicture}
\end{center}
\end{figure}
\newline
Its adjacency matrix $B$ is
$$
B = \begin{pmatrix}
-1 & 1 & 1 & 1\\
1 & -2 & 0 & 0\\
1 & 0 & -3 & 0\\
1 & 0 & 0 & -9\\
\end{pmatrix}\qquad
Tor H_1 (Y) = \intg/3\intg
$$
$$
CS(b) = - (b,B^{-1}b) = \begin{cases}
0\quad mod\quad \intg, & b=(0,0,0,0)\\
\frac{1}{3}\quad mod\quad \intg, & b=(1,0,-1,-6)\\
\end{cases}
$$
We find that $m=3$. Then $\sigma^{18+9} = \lac 1,3,5,7 \rac$, where $K=\lac 1,9\rac$.
\begin{align*}
\hat{Z}_{(1,-1,-1,-1)} & =  q^{71/72} \Psi^{18+9}_{1}(q)\\
\hat{Z}_{(3,-1,-3,-13)} & = -q^{71/72} \Psi^{18+9}_{5}(q).
\end{align*}
\newline

\noindent $Y=M \lb -2 | \frac{1}{2},  \frac{1}{3},  \frac{1}{2} \rb$. Its plumbing graph is
\begin{figure}[h!]
\begin{center}
\begin{tikzpicture}[scale=1]
\centering
\tikzstyle{every node}=[draw,shape=circle]

\draw (0,0) node[circle,fill,inner sep=1pt,label=below:$-2$](){};
\draw (0,0)  -- (1,0) node[circle,fill,inner sep=1pt,label=right:$-2$](){};
\draw (0,0)  -- (-1,0) node[circle,fill,inner sep=1pt,label=left:$-2$](){};
\draw (0,0)  -- (0,1) node[circle,fill,inner sep=1pt,label=right:$-3$](){};

\end{tikzpicture}
\end{center}
\end{figure}
\newline
Its adjacency matrix $B$ is
$$
B = \begin{pmatrix}
-2 & 1 & 1 & 1\\
1 & -2 & 0 & 0\\
1 & 0 & -3 & 0\\
1 & 0 & 0 & -2\\
\end{pmatrix}\qquad
Tor H_1 (Y) = \intg/8\intg
$$
$$
CS(b) = - (b,B^{-1}b) = \begin{cases}
0\quad mod\quad \intg, & b=(0,0,0,0), (1,-1,0,-1)\\
\frac{7}{8}\quad mod\quad \intg, & b= (0,-1,0,0),  (0,0,0,-1)\\
\frac{1}{2}\quad mod\quad \intg, & b= (0,0,-1,0)\\
\end{cases}
$$
We find that $m=6$. Then $\sigma^{18+2} = \lac 1,2,4 \rac$, where $K=\lac 1,2\rac$.
\begin{align*}
\hat{Z}_{(3,-1,-5,-3)} = \hat{Z}_{(3,-3,-5,-1)} & = -\frac{1}{2} q^{-5/12} \Psi^{6+2}_{2}(q)\\
\hat{Z}_{(3,-1,-3,-13)} & = q^{-5/12} \lb 2 q^{1/24} - \Psi^{6+2}_{1}(q) \rb\\
\hat{Z}_{(3,-1,-3,-13)} & = -q^{-5/12} \Psi^{6+2}_{1}(q)\\
\hat{Z}_{(3,-1,-3,-13)} & = q^{-5/12} \Psi^{6+2}_{4}(q).
\end{align*}

\begin{Rmk} The calculations of $\hat{Z}_b$ of Seifert fibered manifolds from the physics approach was done in \cite{HJC}.
\end{Rmk}

\subsection{Line operators}

Line operators in QFTs play an important role. They carry phase structure of the theories. Well known examples of line operators are Wilson and 't Hooft lines. The former informs whether a QFT is in, for example, confining or deconfining phase. In TQFTs, expectation values of line operators yield topological invariants by wrapping knot or links with the operators. In the context of $\hat{Z}$, an insertion of line operators into $\hat{Z}_b$ was first analyzed in \cite{GPPV}. This is natural from the perspective of quantum field theories. Specifically, $3d\hspace{0.1cm} \mathcal{N}=2$ gauge theory $T[M_3;G]$ contain $1/2$-BPS line operators. A knot $K$ in $M_3$ colored by a finite dimensional (irreducible) representation $R$ of $G$ give rises to a line operator $W_{K,R}$.
$$
(K,R) \mapsto W_{K,R} \in \mathcal{C},
$$
where $\mathcal{C}$ is a category of BPS line operators. From M-theory viewpoint, the line operators originate from M2-branes wrapping cotangent bundle of $K$ and located at the origin $O$ of the cigar, $\real \times T^{\ast}K \times O$. We denote the Hilbert space of the $T[M_3;G]$ with $W_{K,R}$ by
\begin{equation}
\mathcal{H}_{T[M_3;G]} \lb D^2 , W_{K,R};b \rb, \qquad b\in Spin^{c}(M_3)/\intg_2.
\end{equation}
It is bigraded carrying homological (R-charge) grading $j$ and q-grading $i$. Thus (16) can be decomposed via the gradings.
$$
\mathcal{H}_{T[M_3;G]} \lb D^2 , W_{K,R};b \rb = \bigoplus\limits_{\substack{i\in \Delta_b + \intg \\ j\in\intg}} H^{i,j}[M_3; W_{K,R}].
$$ 
And the graded Euler characteristics yields the partition function on $S^1 \times D^2$:
\begin{equation}
\chi \lb \mathcal{H}_{T[M_3;G]} \lb D^2 , W_{K,R};b \rb \rb = \hat{Z}_b (M_3,W_{K,R};q) = Z ( S^1 \times D^2,W_{K,R};q) = \sum_{i,j} (-1)^j q^i H^{i,j}[M_3; W_{K,R}].
\end{equation}
From the viewpoint of $\hat{Z}_b (M_3,W_{K,R};q)$, adding $W_{K,R}$ corresponds to inserting the $sl(2)$ character $\chi$ of $R$ into the integrand of (4),
\begin{equation}
\chi_{\lambda (R)} (z) = \frac{z^{\lambda +1} - z^{-\lambda -1}}{z-z^{-1}}.
\end{equation}
In case $M_3$ is a Len space $L(p,1)$, calculations have been carried out in Section 4.3.1 of \cite{GPPV}.
\begin{Rmk} The calculations of $\hat{Z}_b$ of Seifert fibered manifolds containing a knot from the physics approach was done in \cite{HJC2}.
\end{Rmk}

The above situation was generalized to a weakly negative definite plumbed manifold $Y_{\Gamma}$ with multiple Wilson line operators inserted in \cite{CCKPS, CCFFGHP}. Specifically, for $SU(2)$ gauge group, let $\vec{\omega}$ the fundamental weight of $sl(2)$ and insert the line operators $W_{v}$ at vertices $v\in V_{W}$ of $\Gamma$. Then $\hat{Z} (Y_{\Gamma},W_{v};q)$ is given as follows.
\begin{Defn} (\cite{CCKPS}) Consider a weakly negative plumbed manifold $Y_{\Gamma}$, and defects associated to a collection of nodes $V_{W}$ in $\Gamma$, with the highest weight representation with highest weight $\lambda_v \vec{\omega}$. Define the defect $\hat{Z}$ by
\newline
$
\hat{Z} (Y_{\Gamma},W_{v};q) =
$
\begin{equation}
(-1)^{\pi}q^{\frac{3\sigma - \sum_{v\in V} \alpha(v) }{4}} PV \oint \prod_{v\in V } \frac{d z_v}{i2\pi z_v} \lb z_v -\frac{1}{z_v} \rb^{2-\text{deg}(v)} \lb \prod_{v\in V_W} \chi_{\lambda_{v}} (z_v) \rb \Theta^{Y}_{b+ \lac \lambda_v \rac_{v\in V_W}} (q,z),
\end{equation}
where  $\chi_{\lambda_{v}}$ is the $sl(2)$ character (18).
\end{Defn}
We observe that an effect of inserting  the line operators is shifting the $Spin^c$ structures in $\Theta^{Y}$. 

\begin{Prop} (\cite{CCKPS}) The defect $\hat{Z}$ (19) is invariant under the Kirby-Neumann moves in Figure 1 preserving the nodes with $\lambda_{v}\neq 0$. Hence it is a topological invariant of $Y_{\Gamma}$.
\end{Prop}

From viewpoint of quantum modularity, the insertions of the line operators (19) was predicted to realize all components of (11) for a class of plumbed manifolds. This is stated in the following conjecture.
\begin{Conj} (\cite{CCKPS} The modularity conjecture) Two infinite $q$-series are equivalent $f_1 \sim f_2$, if $f_1 = q^{\Delta}f_2 + q^{\Delta^{\prime}}p(q)$, where
$\Delta, \Delta^{\prime} \in\rational$ and $p(q)\in \complex [q^{\pm 1}]$. Consider a Seifert manifold with three singular fibers $M_3$. Define
$$
Span(\hat{Z}(M_3)) := Span_{\complex} \lac \hat{Z}_b (M_3,W_{\bm{\nu}};q)|\, b\in Spin^c (M_3), \bm{\nu}\in N^3 \rac
$$ 
and extend the equivalence between infinite $q$-series to their spans. There exists a Weil representation
$$
\Theta^{(M_3)} =\Theta^{m+K} \quad or \quad \Theta^{(M_3)} =\Theta^{m+K,irr}
$$
for some positive integer $m$ and a subgroup $K\subset Ex_m$ such that the following is true.
\begin{enumerate}
	\item When $M_3$ is negative definite, $Span(\hat{Z}(M_3))$ is equivalent to $Spin_{\complex}\lac \tilde{\theta}_{r}^{(M_3)} | r\in \intg/2m \rac$
	
	\item When $M_3$ is positive definite, there is a $SL(2,\intg)$ vector-valued (mixed) mock modular form $f^{(M_3)}= (f^{(M_3)}_{r})$ transforming in the dual representation of $\Theta^{(M_3)}$ such that $Span(\hat{Z}(M_3))$ is equivalent to $Spin_{\complex}\lac f^{(M_3)}_{r} | r\in \intg/2m \rac$.	
\end{enumerate}
\end{Conj}
This conjecture was proved for the Brieskorn spheres.
\begin{Thm} (\cite{CCKPS}) The Conjecture 2.18 is true for Brieskorn spheres $M_3 =\Sigma(p_1,p_2,p_3)$. More precisely, we have
$$
\hat{Z}(\Sigma(p_1,p_2,p_3), W_{\bm{\nu}};q) = c q^{\Delta} \tilde{\theta}^{m+K}_{r_{\bm{\nu}}} + p(q),\quad r_{\bm{\nu}}= m - \sum_{i} (1+\nu_i)\bar{p}_i,
$$
where $\bar{p}_i = m/p_i , p(q)$ is a (possibly vanishing) polynomial and $c\in \complex$.
\end{Thm}
\begin{Rmk} The modularity data of $\Sigma(p_1,p_2,p_3)$ is stated in (13).
\end{Rmk}
We illustrate the above conjecture via examples~\cite{CCKPS}.
\newline

\noindent\textbf{Example}
\noindent $M_3 = \Sigma(2,3,7)$: The modular data $m$ and $K$ are
$$
m=42, \quad K=\lac 1,6,14,21\rac,
$$
and $\sigma^{m+K} = \lac 1,5,11\rac$.
$$
\hat{Z} (M_3;q) \sim \tilde{\theta}^{m+K}_1.
$$
In the absence of a line operator, we only have one element in $\sigma^{m+K}$. The other two can be realized via a line operator insertion.
\begin{align*}
\hat{Z} (M_3, W_{0,0,1};q) & \sim  \tilde{\theta}^{m+K}_5\\
\hat{Z} (M_3, W_{0,0,2};q) &  \sim  \tilde{\theta}^{m+K}_{11}.
\end{align*}
\newline

\noindent $M_3 = M\lb -2 | \frac{1}{2},\frac{2}{3},\frac{2}{3}\rb$. Its $H_1 (M_3) =\intg/3\intg$.
$$
m=6, \quad K=\lac 1,3\rac,
$$
and $\sigma^{m+K} = \lac 1,3\rac$.
\begin{align*}
\hat{Z}_0 (M_3 ;q) & \sim  \tilde{\theta}^{m+K}_1\\
\hat{Z}_{(1, 1,−2, 1, 0,−1)} (M_3 ;q) & \sim  \tilde{\theta}^{m+K}_3\\
\end{align*}
They span all the components. In this case, the line defect insertions result in
\begin{align*}
\hat{Z} (M_3 , W_{(0,0,1)} ;q) &  \sim  \tilde{\theta}^{m+K}_3\\
\hat{Z} (M_3 , W_{(0,0,2)};q) &  \sim  \tilde{\theta}^{m+K}_{1}.
\end{align*}

\subsection{Effective central charge}

The intergrality of coefficients of $\hat{Z}_b$ is a core feature of its topological invariance and its clue to existence of the deeper algebraic structure. From the physics perspective, coefficients of (supersymmetric) partition function  or (superconformal) index of a (supersymmetric) QFT reflect the dimensions of sectors of  BPS Hilbert spaces of the theory. This is in turn tied to the central charge of the theory via counting of dynamical degrees of freedom. In \cite{GJ}, the coefficients $a_n \in \intg$ of (5) were analyzed in the context of strongly coupled $3d\hspace{0.1cm}\mathcal{N}=2$ superconformal field theories (SCFTs). Specifically, it was shown that $a_n$ has a particular growth behavior as a function of $n$ and the specifics of the behavior was encoded via an analogue of central charge of the theories, which was called effective central charge $c_{eff}$. We begin with the following prediction about the BPS states of $3d\hspace{0.1cm} \mathcal{N}=2$ SCFTs.
\begin{Conj}(\cite{GJ}) In every $3d\hspace{0.1cm} \mathcal{N}=2$ SCFT, the spectrum of supersymmetric (BPS) states obeys
\begin{equation}
a_n \sim e^{2\pi \sqrt{\frac{c_{eff} n}{6}}}.
\end{equation}
 In other words, coefficients $a_n$ of the superconformal index or, equivalently, $S^2 \times_q S^1$ partition function,
\begin{equation}
I(q) = Tr_{H_{S^2}} \lsb (-1)^F q^{R/2 + J_3} \rsb = Z(S^2 \times_q S^1) = \sum a_n q^n
\end{equation}
enjoy (20).
\end{Conj}

\begin{Defn}(\cite{GJ}) Assuming Conjecture 2.21, to any $3d\hspace{0.1cm} \mathcal{N}=2$ SCFT we associate a quantity $c_{eff}$ defined via the asymptotic behavior of superconformal index (21):
\begin{equation}
c_{eff} := \frac{3}{2\pi^2} \lim_{n\rightarrow \infty} \frac{\lb log(|a_n |)\rb^2}{n}.
\end{equation}
\end{Defn}
It is expected that (22) measures the number of degrees of freedom of $3d\hspace{0.1cm} \mathcal{N}=2$ SCFTs.
\newline

Evidence for the above conjectures was provided for non-negative definite Brieskorn spheres $-Y= -\Sigma(s,t,rst \pm 1)\, (gcd(s,t)=1),\, r\in \intg_{+}$ in \cite{GJ}. The coefficients $a_n$ in (21) grow as
$$
a_n (r,s,t) \sim exp \lsb \sqrt{16\pi^2 \lb \frac{m^2}{4st(rst\pm 1)} - l \rb n} \rsb.
$$
From numerical analysis, it was concluded that $l=0$ for $-Y$. And values of $m$ are estimated to be the following.
\begin{center}
\begin{tabular}{ |c|c|c|c| } 
\hline
$t \downarrow \quad s \rightarrow$  & 2 & 3 & 4 \\ 
 \hline
$4$    &   & 3.90 &    \\ 
 \hline
$5$     & 2.72 & 5.01 & 6.69   \\ 
 \hline
$7$    & 4.35 & 7.10 & 9.54   \\ 
 \hline
$8$    & & 8.16 &    \\ 
 \hline
$9$    & 5.88 &  &   12.38 \\ 
 \hline
$10$    & & 10.27 &    \\ 
 \hline
$11$    & 7.36 & 11.33 & 15.21   \\ 
 \hline
$13$    & 8.22 & 13.45 & 18.02   \\ 
\hline
\end{tabular}
\end{center}
Hence, the formula of $c_{eff}$ for $-Y$ is given by
$$
c_{eff} = 2+ \frac{24m^2}{4st(rst\pm 1)}.
$$
\begin{Rmk} We will describe $\hat{Z}_b$ for orientation reversed manifolds $-Y$ in Section 2.7.
\end{Rmk}
\begin{Rmk} Further investigations on $c_{eff}$ were carried out recently in \cite{HJNP, ACDGO}~\footnote{The papers appeared when the review article was in preparation.}.
\end{Rmk}

\subsection{Relations to other invariants}

\begin{figure}[h!]
\centering
\includegraphics[scale=0.5]{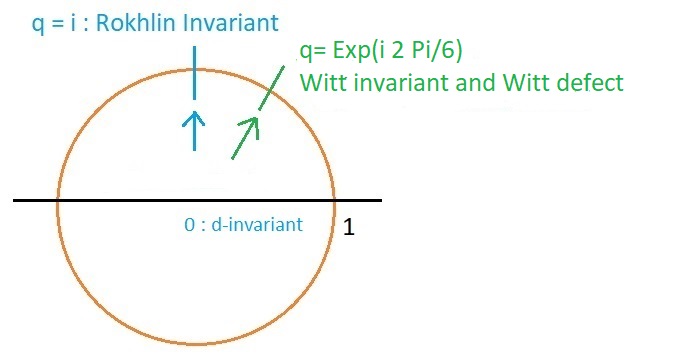}
\caption{Topological invariants at the fourth and the sixth roots of unity from the limits of $\hat{Z}_b$.}
\end{figure}

A connection between the quantum invariant $\hat{Z}_b (q)$ and topological invariants were first found in \cite{GPP}. The authors investigated the spin refined version of the WRT invariant at the fourth root of unity and elucidated that the corresponding $\hat{Z}$'s are related to the Rokhlin invariant $\mu(Y,s)$ and the $d$-invariant (or the correction term) of a certain version of the Heegaard Floer homology for several classes of 3-manifolds.
\newline

Among a variety of topological invariants, an interesting one is cobordism invariant. It establishes a relation between n-dimensional manifolds $M^{n}_1 , M^{n}_2$ via cobordism $(n+1)$-dimensional manifold $N^{n+1} (\ptl N^{n+1} = M^{n}_1 \cup -M^{n}_2$). This includes whether $M^{n}$ can bound a $N^{n+1}$. This feature is informed by the cobordism groups $\Omega (n)$. If $\Omega (n)=0$, then this implies $M^{n}$ can bound $M^{n+1}$. If $M^{n}$ is equipped with Spin structure, then the relevant cobordism groups are Spin cobordism groups $\Omega^{Spin} (n)$. When $n=3$, it vanishes. Hence spin $M^3$ bounds a (topological or smooth) spin four manifold $M^4$. The spin structure of the latter originates from the former by an extension. Rokhlin proved that a $M^{3}$ invariant depending on its Spin structure $s$ is related to the signature of $M^4$ up to mod 16~\cite{Rok}:
$$
\mu(M^3 ,s) = \sigma(M^4)\qquad \text{mod}\, 16.
$$
An implication is that if $M^4$ is smooth, then $\mu(M^3 ,s)=0$.
\newline

A precise relation between $\hat{Z}$ and $\mu(Y,s)$ was shown in \cite{GPP}:
$$
e^{-i 2\pi \frac{3\mu(Y,s)}{16}} = \sum_{b} c^{Rokhlin}_{s,b} \hat{Z}_{b}[Y;q] \bigg|_{q \rarw i}\qquad s \in Spin(Y),\quad b \in  Spin^c (Y).
$$
Furthermore, the overall exponent $\Delta_b$ in $\hat{Z}$ is related to $d$-invariant is related by
$$
\Delta_b (Y,b) = \frac{1}{2} - d(Y,b) \qquad \text{mod}\, 1.
$$ 
In case of $Y=\intg HS^3$,
$$
c^{Rokhlin} = \frac{1}{i\sqrt{8}}.
$$
In case of $Y=\rational HS^3$ (i.e $H_1(Y)= \intg/p\intg$),
\begin{align*}
p & = \text{odd},\qquad c^{Rokhlin}_{b}  = \frac{1}{8} \sum_{n=0}^7 e^{-i\pi \frac{(np-2b)^2 +2p}{8p}}\\
p & = \text{even},\qquad c^{Rokhlin}_{s,b} = \frac{1}{4} e^{-i\pi \frac{2(b+sp/2)^2 +p}{4p}} \lb 1+ (-1)^b e^{(s-1)\frac{i\pi p}{2}} \rb.
\end{align*} 
In general, $Y$ whose first Betti number $b_1 (Y)=0$, then $c^{Rokhlin}$ is given by
\begin{align*}
c^{Rokhlin}_{s,\sigma(s,b)} & = \frac{1}{i\sqrt{8|H_1 (Y)|}} \sum_{a\in H_1 (Y)} e^{-i2\pi lk(a,a) - i2\pi lk(a,b)},\\
\sigma & : Spin(Y) \times H_1 (Y,\intg) \rightarrow  Spin^c (Y)\\
lk & : Tor H_1 (Y,\intg) \times Tor H_1 (Y,\intg) \rightarrow \rational / \intg,
\end{align*} 
where $lk$ is the linking form.
\newline

The invariant $\Delta (Y,b)$ was further analyzed for negative definite plumbed manifolds $Y_{\Gamma}$ in \cite{HNS}. It was found that $\Delta_b (Y,b)$ is related to a topological invariant $\gamma (Y) := k^2 + s$, where $s$ is the number of vertices of the plumbing graph and $k$ is the characteristic vector. The latter is an element in $H^2 (W)$, where $W$ is a four manifold bounded by $Y$. In case of negative definite plumbed manifolds,  $\gamma (Y)$ can be expressed in terms of data of plumbing graph.
\begin{Prop} (\cite{NN}) Let $Y_{\Gamma}$ be a negative definite plumbed manifold, which is a rational homology sphere. Then
$$
\gamma (Y)  = 3s+ Tr(B) + 2+ (2u- \delta)^2,
$$ 
where $u=(1,\cdots, 1)$ and $\delta = (\delta_v)_{v\in V}$ is the degree vector ($V$ is the set of vertices of $Y_{\Gamma}$).
\end{Prop}
We state the relation between $\Delta (Y,b)$ and $\gamma (Y)$.
\begin{Thm} (\cite{HNS}) Let $Y=M (b_0 | (a_1, w_1),\cdots, (a_n, w_n) $ be a Seifert fibered manifold with $n$ singular fibers associated a negative definite plumbing graph. Let \textit{can} be the canonical $Spin^c$ structure of $Y$. Then $\Delta_{can}$ satisfies
$$
\Delta_{can} = -\frac{\gamma(Y)}{4} + \frac{1}{2}.
$$
If $Y$ is not a lens space, then $\Delta_{can}$ is minimal among all $\Delta_{b},\, b\in Spin^c (Y)$.
\end{Thm} 

\indent Another connection between the quantum invariant $\hat{Z}_b$ and topological invariants were found in \cite{C2}. Specifically, a new relation between the Witt invariant $w(Y)$, Witt defect $\text{def}_{3}(\Theta)$ and $\hat{Z}_b$ of $Y$ from a certain refinement of the WRT invariant at the sixth root of unity was established. In \cite{KMZ}, the $SU(2)$ WRT invariant at the sixth root of unity for a closed oriented 3-manifold was investigated. It was shown that the WRT invariant is a sum of the invariants of the manifold equipped with a 1-dimensional mod 2 cohomology class $\Theta$:
$$
\tau_{6}[Y] = \sum_{\Theta \in H^{1}(Y; \intg/ 2\intg)} \tau_{6}[Y, \Theta].
$$
Furthermore, $\tau_{6}[Y, \Theta]$ can be expressed in terms of $w(Y)$ and $\text{def}_{3}(\Theta)$,
\begin{equation}
\tau_{6}[Y, \Theta] = i^{-w(Y) + 2\Theta^3 + \text{def}_{3}(\Theta)} \sqrt{3}^{\epsilon(\Theta) + d(Y_{\Theta}) - d(Y)}
\end{equation}
\newline
$$
d(Y) = rk H^{1}(Y; \intg/ 3\intg), \qquad  d(Y_{\Theta}) = rk H^{1}(Y_{\Theta}; \intg/ 3\intg), \qquad 2\Theta^3 \in \intg / 4\intg
$$
$$
w(Y): \text{mod 3 Witt invariant of}\, Y \quad (cf. (4)) 
$$
$$
\text{def}_{3}(\Theta) : \text{mod 3 Witt defect of the double cover manifold}\, Y_{\Theta} \rarw Y \quad (cf. (5))
$$
$$
\epsilon(\Theta)=   \begin{cases}
0,\quad \Theta =0 \\
1,\quad \Theta \neq 0\\
\end{cases}
$$
\newline

Let us first review the Witt invariant and defect of 3-manifolds defined in \cite{KMZ}. Their formulation takes place in four dimension. Let $Y$ be a closed oriented 3-manifold. By the vanishing of its oriented cobordism group $\Omega(Y)=0$, $Y$ bounds a compact oriented 4-manifold $X$ whose intersection form is denoted by $\tilde{Q}_X$. Its signature is denoted by $\sigma(X)$. We next diagonalize $\tilde{Q}_X$ in $\intg_3$-coefficient ring, obtaining $0, \pm 1$ as its diagonal entries. We denote it by $Q_X$. Then we let $w(X)$ to be its trace Tr $Q_X$. The mod 3 Witt invariant of $Y$ is defined as
\begin{equation}
w(Y) : = \sigma(X) -  w(X)\quad \text{mod}\, 4.
\end{equation}
$w(Y)$ is independent of $X$. Since we deal with a compact 4-manifold with a boundary, we would like to detect an effect of the boundary. This leads to the notion of the Witt defect. Specifically, we consider a cyclic n-fold cover manifold $\tilde{Y} \rarw Y$. By the result of \cite{CG}, this covering manifold extends to a cyclic branched cover $\tilde{X} \rarw X$ branched along a closed surface $F$ in $X$. We let $Q_{\tilde{X}}$ be an intersection form of $\tilde{X}$ in $\intg_3$ coefficient. The mod 3 Witt defect of $\tilde{Y} \rarw Y$ is defined as
$$
\text{def}_{3}(\tilde{Y} \rarw Y) := n w(X) - w(\tilde{X}) - \frac{n^2 -1 }{3n} F \cdot F \quad \text{mod}\, 4,
$$
where n divides $F \cdot F$. The specific Witt defect that is relevant in our context is a double cover 3-manifold equipped with a cohomological class $\Theta \in H^{1}(Y ; \intg/2\intg)$:
\begin{equation}
\text{def}_{3}(Y_{\Theta} \rarw Y) = 2 w(X) - w(\tilde{X}) - \frac{1}{2} F \cdot F \quad \text{mod}\, 4.
\end{equation}
We abbreviate the above defect as $\text{def}_{3}(\Theta)$. Due to the presence of the boundary, the difference between the first two terms in (25) is not necessarily zero. Note that $w(Y)$ and $\text{def}_{3}(Y_{\Theta} \rarw Y)$ taking value in $\intg/4\intg$ follows from the fact that the Witt ring $W(R)$ of $R=\intg/3\intg$ is $\intg/4\intg$~\cite{MH}.
\newline

The Witt invariant $w(Y)$, Witt defect $\text{def}_{3}(\Theta)$ are geometrically defined on the level of 4-manifolds, thus they also posses cobordism characteristic.
$$
 i^{-w(Y) + 2\Theta^3 +  \text{def}_{3}(\Theta)} \sqrt{3}^{\epsilon(\Theta) + d(Y_{\Theta}) - d(Y)} =  \sum_{b\in Spin^c (Y) / \intg_2} c^{Witt}_{\Theta b} \hat{Z}_{b} (q)\bigg|_{q \rarw e^{\frac{i2\pi}{6}}}\qquad \Theta \in H^{1}(Y; \intg / 2\intg).
$$
For rational homology spheres $Y$ ($H_1 (Y; \intg) = \intg / p\intg $), there are two different cases. The first case is when $p=$ odd, 
\begin{equation}
c^{Witt}_{t}= \frac{e^{-i\pi/4}}{4\sqrt{3}} \sum_{r=0}^{5} e^{-\frac{i\pi}{12p} (2pr-2t+p)^2},
\end{equation}
where $t=0,\cdots, p-1$. When $p=$even,
\begin{equation}
c^{Witt}_{w t}= \frac{e^{-i\pi/4}}{2\sqrt{3}}e^{-\frac{i\pi}{3p} (t+\frac{p}{2}(w+1))^2} \lb 1 + e^{\frac{i\pi}{3} (pw+2t)} + e^{\frac{i2\pi}{3} (pw+2t-p)} \rb ,
\end{equation}
where $w=0,1$ and $t=0,\cdots, p-1$~\footnote{We used the fact that $Spin^c (Y)$ is affinely isomorphic to $H_1 (Y; \intg)$}. This new relation not only enriches the conceptual aspects of the invariants, it provides a new method of computing the Witt invariant and Witt defect directly in three dimension as well.
\newline

\noindent As examples, we list the Witt invariants for $L(p,1)$. 
\begin{center}
\begin{tabular}{ |c|c|c| } 
\hline
$-L(p ,1 )$  & $w(Y) \in \intg /4\intg $ & $d(Y) \in \intg $ \\ 
 \hline
$-L(3,1)$    & 3   & 1   \\ 
 \hline
$-L(5,1)$    & 2  & 0 \\ 
 \hline
$-L(7,1)$    & 0  & 0 \\ 
\hline
\end{tabular}
\end{center}
\begin{center}
\begin{tabular}{ |c|c|c|c|c|c| } 
\hline
$-L(p ,1 )$  & $w(Y) \in \intg /4\intg $ & $d(Y) \in \intg$  & $d(Y_1) \in \intg$ & $\text{def}_{3}(1)  \in \intg /4\intg $ & $ 2(1^3) \in \intg /4\intg $   \\ 
\hline
$-L(2,1)$  &  2  & 0 & 0 & 0 & 2 \\ 
\hline
$-L(4,1)$  &  0  & 0 & 0 & 3 & 0\\ 
\hline
$-L(6,1)$  &  3  & 1 & 1 & 3 & 2 \\ 
\hline
$-L(8,1)$  &  2  & 0 & 0 & 3 & 0 \\
\hline 
$-L(10,1)$ & 0 & 0 & 0 & 2 & 2 \\ 
\hline
$-L(12,1)$ & 3 & 1 & 1 & 2 & 0 \\ 
\hline
$-L(14,1)$ & 2 & 0 & 0 & 2 & 2  \\ 
\hline
$-L(16,1)$ & 0 & 0 & 0 & 1 & 0  \\ 
\hline
$-L(18,1)$ & 3 & 1 & 1 & 3 & 0  \\
\hline 
$-L(20,1)$ & 2 & 0 & 0 & 1 & 0  \\ 
\hline
$-L(22,1)$ & 0 & 0 & 0 & 0 & 2  \\ 
\hline
$-L(24,1)$ & 3 & 1 & 1 & 2 & 2  \\ 
\hline
\end{tabular}
\end{center}

\subsection{Orientation reversal}

Under orientation reversal of a closed oriented 3-manifold $Y \rightarrow -Y$, the $sl(2)$ WRT invariants of $-Y$ at level $k$ is
$$
WRT(-Y;-k)= WRT(Y;k)^{\ast}.
$$
Since the WRT invariant is a complex number, the orientation reversal amounts to applying the complex conjugation, equivalently, sending $k \rightarrow -k$. The latter implies $q \rightarrow 1/q$. In case a topological invariant is a series, for instance, $\hat{Z}_b (Y) (q)$,  orientation reversal of $Y$ translates to a nontrivial operation. Naively sending $q \rightarrow 1/q$ does not lead to a correct series for $\hat{Z}_b (-Y) (q)$. As a consequence, a major challenge in the development of $\hat{Z}_b (Y) (q)$ is finding a formula for positive definite plumbed manifolds~\footnote{For lens spaces $L(m,n)$, their $\hat{Z}_b (L(m,n)) (q)$ are monomial. Hence $\hat{Z}_b (-L(m,n)) (q)$ can be obtained by $q \rightarrow 1/q$.}. In case of $\hat{Z}(-Y)_b (q)$, it is a $q$ power series defined on the outside the unit disk in the complex plane $(|q|>1)$. From the viewpoint of quantum modularity in Section 2.3, $\hat{Z}_b (Y) (q)$ of certain manifolds can be expressed in terms of a linear combinations of the false theta functions. In \cite{CCFGH}, it was shown that $\hat{Z}_b (-Y) (q)$ can be expressed in terms of the mock theta functions. And there exists a false-mock pair for between $\hat{Z}_b (Y) (q)$ and $\hat{Z}_b (-Y) (q)$. 
\newline

In the physics context, let us recall that $\hat{Z}_b (-Y) (q)$ appeared in Section 2.2 and 2.5. It is a necessary ingredient in analyzing the superconformal index of $3d\hspace{0.1cm} \mathcal{N}=2$ SCFTs, and the effective central charge. 
\newline
 
Although there are available approaches to find $\hat{Z}_b (-Y) (q)$, they have limited applicability or difficulty to implement in practice. We summarize the approaches.

\begin{enumerate}
	\item $q$-hypergeometric series: It was shown in \cite{CCFGH} that $\hat{Z}_b (Y) (q)$ of particular Seifert fibered manifolds with three singular fibers, for example, $\Sigma(2,3,5)$ and $\Sigma(2,3,7)$ can be expressed in terms of $q$-hypergeometric series. It is a rational function, which allows us to apply $q \rightarrow 1/q$ to obtain $\hat{Z}_b (-Y) (q)$ straightforwardly. 
	
	\item Rademacher sums: This method is systematic and sophisticated~\cite{CCFGH}. Obtaining $\hat{Z}_b (-Y) (q)$ involves finding a certain function in the lower half of the complex plane, which is an image of mock modular form.
		
	\item Resurgence method: This method utilizes quantum modular property of $\hat{Z}_b (Y;q)$~\cite{CDGG}~\footnote{Resurgence method is a technique that enables to analyze strongly coupled QFTs. It was applied to the complex Chern-Simons theory in \cite{GMP}.}. The method aims to find a dual $\Psi_{m,r}(q)^{\vee} $ of a false theta functions $\Psi_{m,a}(q)$:
	$$
	\Psi_{m,a}(q) \longleftrightarrow \Psi_{m,a}(q)^{\vee} 
	$$
		A starting point for obtaining $\Psi_{m,a}(q)^{\vee}$ is the Borel-Mordell integral and its unique decomposition,
\begin{equation}
	J_{(p,a)}(\hbar) := \frac{1}{-\hbar} \int_{0}^{\infty} du e^{pu^2/\hbar} \frac{sinh((p-a)u)}{sinh(pu)},\qquad \hbar <0.
\end{equation}
	$$
	\sqrt{\frac{4p(-\hbar)}{\pi}} J_{(p,a)}(\hbar) = q^{\Delta_a} F_a (q) + i \sqrt{\frac{\pi}{\hbar}} \sum_{b=1}^{Floor\lsb \frac{p}{2} \rsb} S_{ab} \tilde{q}^{\tilde{\Delta}_b} W_b (\tilde{q}), \qquad q=e^{\hbar},\quad \tilde{q}=e^{\frac{\pi^2}{\hbar}},
	$$
  where $F_a (q)$ is a $q$ series and $W_b$ is a $\tilde{q}$ series. Using (28) and the numerical analysis, $F_a (q)$ and $W_b (\tilde{q})$ are determined~\footnote{Analytic continuation of (28) into $\hbar >0$ regime is given in Section 4.6 of \cite{CDGG}.}, they include the desired mock theta functions $\Psi_{m,a}(q)^{\vee}$.	
\begin{Rmk} In the above, $p=m$ and $a=r$ from Section 2.3.  
\end{Rmk}

\item Indefinite theta functions: Instead of using the negative definite lattice for the theta function in (4), an application of indefinite lattice theta functions was introduced in \cite{CFS}. Specifically, it used an indefinite lattice theta function together with a regulator. In this approach, there is a choice of a one sided cone when summing over lattice vectors. This idea was further pursued and refined in \cite{P3}, which used a double cone.
\end{enumerate}

\section{Series invariant for link complements}

\subsection{A two variable series for knots}

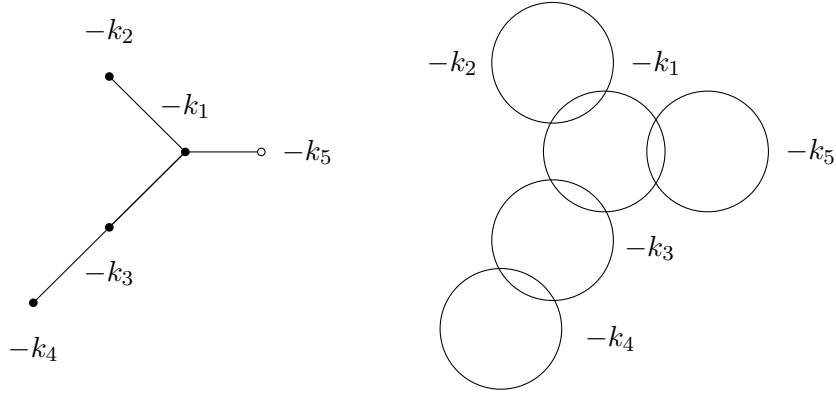
\begin{figure}[t!]
\begin{center}
\begin{tikzpicture}[scale=1]
\centering
\tikzstyle{every node}=[draw,shape=circle]

\draw (0,0) node[circle,fill,inner sep=1pt,label=above:$-k_1$](){};
\draw (0,0)  -- (1,0) node[circle,fill=white,inner sep=1pt,label=right:$-k_5$](){};
\draw (0,0)  -- (-1,1) node[circle,fill,inner sep=1pt,label=above:$-k_2$](){};
\draw (0,0)  -- (-1,-1) node[circle,fill,inner sep=1pt,label=below:$-k_3$](){};
\draw (0,0)  -- (-2,-2) node[circle,fill,inner sep=1pt,label=below:$-k_4$](){};

\end{tikzpicture}
\qquad
\begin{tikzpicture}[scale=0.8]
\centering

\draw (0,0) circle (1cm);
\draw (-0.85,1.47) circle (1cm);
\draw (-0.85,-1.47) circle (1cm);
\draw (-1.7,-2.94) circle (1cm);
\draw (1.7,0) circle (1cm);

\node at (0.85,1.47) {$-k_1$};
\node at (-2.5,1.47) {$-k_2$};
\node at (0.75,-1.6) {$-k_3$};
\node at (0.1,-3.1) {$-k_4$};
\node at (3.4,0) {$-k_5$};

\end{tikzpicture}

\end{center}
\caption{A plumbing graph $\Gamma_K$ of a knot $K \subset S^3$ (left) and corresponding surgery link $L(\Gamma_K)$. The linking between two link components is the Hopf link. This link diagram can be transformed into a knot diagram through the Kirby moves.}
\end{figure}

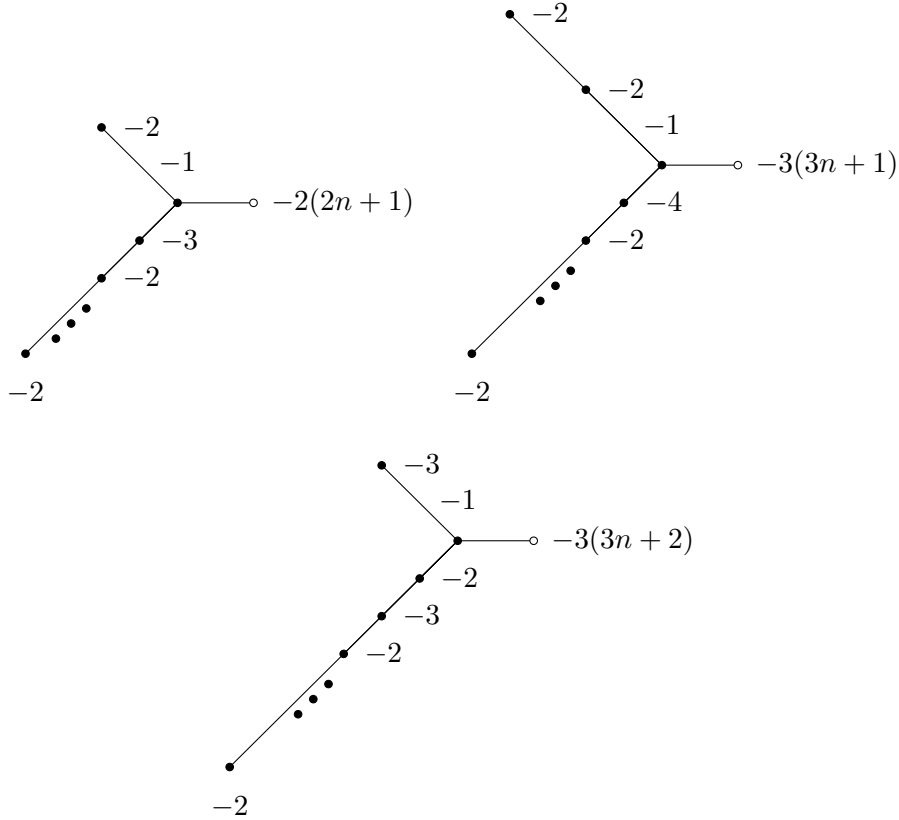
\begin{figure}[t!]
\begin{center}
\begin{tikzpicture}[scale=1]
\centering
\tikzstyle{every node}=[draw,shape=circle]

\draw (0,0)  node[circle,fill,inner sep=1pt,label=above:$-1$](){}; 
\draw (0,0) -- (1,0) node[circle,fill=white,inner sep=1pt,label=right:$-2(2n+1)$](){};
\draw (0,0)  -- (-1,1) node[circle,fill,inner sep=1pt,label=right:$-2$](-2){};

\draw (0,0) -- (-0.5,-0.5) node[circle,fill,inner sep=1pt,label=right:$-3$](-3){};
\draw (0,0)  -- (-1,-1) node[circle,fill,inner sep=1pt,label=right:$-2$](-2){};
\draw (0,0) -- (-2,-2) node[circle,fill,inner sep=1pt,label=below:$-2$](){};

\draw (-1.2,-1.4) node[circle,fill,inner sep=1pt,label=above:$$](){};
\draw (-1.4,-1.6) node[circle,fill,inner sep=1pt,label=above:$$](){};
\draw (-1.6,-1.8) node[circle,fill,inner sep=1pt,label=above:$$](){};

\end{tikzpicture}
\begin{tikzpicture}[scale=1]
\tikzstyle{every node}=[draw,shape=circle]

\draw (0,0) node[circle,fill,inner sep=1pt,label=above:$-1$](){};
\draw (0,0) -- (1,0) node[circle,fill=white,inner sep=1pt,label=right	:$-3(3n+1)$](){};
\draw (0,0)  -- (-1,1) node[circle,fill,inner sep=1pt,label=right:$-2$](-2){};
\draw (0,0)  -- (-2,2) node[circle,fill,inner sep=1pt,label=right:$-2$](-2){};

\draw (0,0) -- (-0.5,-0.5) node[circle,fill,inner sep=1pt,label=right:$-4$](-4){};
\draw (0,0)  -- (-1,-1) node[circle,fill,inner sep=1pt,label=right:$-2$](-2){};
\draw (0,0) -- (-2.5,-2.5) node[circle,fill,inner sep=1pt,label=below:$-2$](-2){};

\draw (-1.2,-1.4) node[circle,fill,inner sep=1pt,label=above:$$](){};
\draw (-1.4,-1.6) node[circle,fill,inner sep=1pt,label=above:$$](){};
\draw (-1.6,-1.8) node[circle,fill,inner sep=1pt,label=above:$$](){};

\end{tikzpicture}
\begin{tikzpicture}[scale=1]
\centering
\tikzstyle{every node}=[draw,shape=circle]

\draw (0,0) node[circle,fill,inner sep=1pt,label=above:$-1$](-1){}; 
\draw (0,0) -- (1,0) node[circle,fill=white,inner sep=1pt,label=right:$-3(3n+2)$](){};
\draw (0,0)  -- (-1,1) node[circle,fill,inner sep=1pt,label=right:$-3$](-3){};

\draw (0,0) -- (-0.5,-0.5) node[circle,fill,inner sep=1pt,label=right:$-2$](-2){};
\draw (0,0)  -- (-1,-1) node[circle,fill,inner sep=1pt,label=right:$-3$](-3){};
\draw (0,0)  -- (-1.5,-1.5) node[circle,fill,inner sep=1pt,label=right:$-2$](-2){};
\draw (0,0) -- (-3,-3) node[circle,fill,inner sep=1pt,label=below:$-2$](-2){};

\draw (-1.7,-1.9) node[circle,fill,inner sep=1pt,label=above:$$](){};
\draw (-1.9,-2.1) node[circle,fill,inner sep=1pt,label=above:$$](){};
\draw (-2.1,-2.3) node[circle,fill,inner sep=1pt,label=above:$$](){};

\end{tikzpicture}
\end{center}
\caption{Plumbing graphs of $T(2,2n+1)$ (left), $T(3,3n+1)$ (right) and, $T(3,3n+2)$ (bottom). The ellipsis indicates intermediate vertices with weight $-2$ along the legs. Total number of $-2$ vertices in succession on the leg is $n-1$ for $T(2,2n+1), T(3,3n+1)$ and $T(3,3n+2)$.}
\end{figure}

Motivated by (4), a multi-variable series for complements of plumbed knots was defined in \cite{GM}. We begin by reviewing plumbed knots.\\
\indent Plumbed knot complements, more generally, plumbed 3-manifolds with a torus boundary, are represented by a weighted graph $\Gamma_K$ with one distinguished vertex $v_{\ast}$~\cite{GM}. This vertex represent the torus boundary. We are interested in the case when degree of $v_{\ast}$ is one. From the viewpoint of the surgery link $L(\Gamma_K)$ described above, an unknot corresponding to $v_{\ast}$ acts as a spectator during the surgery operation. Furthermore, removing $v_{\ast}$ and the edge connecting it to $\Gamma_K$ represent an ambient plumbed 3-manifold $Y(\hat{\Gamma})$. 
\newline
\indent Additional data describing a knot is framing that takes values in $\intg$. Roughly speaking, this value characterizes twisting of a longitude of the knot around the knot. This information is captured by weight $k_{v_{\ast}}$ of $v_{\ast}$. This is called \textit{graph framing}. Therefore, complement of a plumbed knot in $Y(\hat{\Gamma})$ is specified by $(\Gamma_K , v_{\ast})$. A simple example is shown in Figure 3. The Neumann moves in Figure 1 also apply to plumbing graphs of knots, except the blow up/down move on $v_{\ast}$.
\newline

We review the method for obtaining plumbing graphs of torus knots in \cite{GM}. We consider torus knots $T(s,t) \subset S^3$ where $gcd(s,t) =1,\, 2 \leq s < t$. Torus knots are examples of algebraic knots. Hence they, more precisely, their complements admit plumbing graph presentations. The graphs consist of one multivalency vertex having degree $3$ and weight $-1$ and three legs attached to the vertex. One of the legs has an open vertex of degree 1 called distinguished vertex representing a torus boundary of the knot complement. To find vertices and weights on the other legs, we solve 
$$
\frac{t^{\prime}}{t} + \frac{s^{\prime}}{s} = 1- \frac{1}{st}
$$
for unique integers $t^{\prime} \in (0,t)$ and $s^{\prime} \in (0,s)$ satisfying
$$
st^{\prime} \equiv -1 \, (\text{mod}\, t) \qquad ts^{\prime} \equiv -1 \, (\text{mod}\, s).
$$
Then we expand $-t/t^{\prime}$ and $-s/s^{\prime}$ in continued fractions in Section 3.1. Each of them forms a leg with weights attached to the central vertex. The weight of the distinguished vertex is given by $-st$~\footnote{This value corresponds to 0-framed torus knots.}. Example of plumbing graphs are shown in Figure 4. 
\newline

For complements of plumbed knots that are (weakly) negative definite $(\Gamma_K , v_{\ast})$, a three variable series was defined~\cite{GM}:
\begin{align}
\begin{split}
\hat{Z}_b (\Gamma_K ;z,n,q) & = (-1)^{\pi} \lb z- \frac{1}{z} \rb^{1-\text{deg}(v_{\ast})} q^{\frac{3\sigma - TrB}{4}} \prod_{\substack{v\in Vert \\ v\neq v_{\ast}}}\oint_{|z_v|=1} \frac{dz_v}{i2\pi z_v} \lb z_v- \frac{1}{z_v} \rb^{2-\text{deg}(v)} \Theta^{-Y_K}_b (\vec{z},q),\\
\Theta^{-Y_K}_b & = \sum_{\vec{l}} q^{-\frac{(\vec{l},B^{-1}\vec{l})}{4}} \prod_{v\in Vert} z_{v}^{l_v},
\end{split}
\end{align}
where $Vert$ denotes a set of vertices of $\Gamma_K$ and $B$ is the linking matrix of  $L(\Gamma_K)$. We note that there is no integration over $v_{\ast}$ in (29).
\newline

Using the properties of the plumbed knot complements, it was shown that (29) reduces to an independent two variable series denoted by $F_K (x,q)$. It turns out that $F_K (x,q)$ has the following general form for all knots in $\intg HS^3$.
\begin{equation}
F_K (x,q) = \sum_{\substack {m =1 \\ \text{odd} } }^{\infty} f_m (K;q) \lb x^{m/2} - x^{-m/2} \rb \quad \in  \frac{1}{2^c} q^{\Delta} \intg [q^{-1}, q]][[x^{1/2}, x^{-1/2}]]
\end{equation}
where $c\in\intg_{\geq 0}, \Delta \in\rational$ and $x=z^2$.
\begin{Rmk} The variable $x$ in (30) counts with the relative $Spin^c$ structures of the knot complement.
\end{Rmk}
\begin{Rmk} Generally, the coefficient functions $\lac f_m (K;q) \rac$ are a Laurent power series $\intg[q^{-1},q]]$; knots whose Alexander polynomials are non-monic have this property. In case of fibered knots, they are Laurent polynomials.
\end{Rmk}

\subsection{Large color R-matrix }

\begin{figure}[h!]
\centering
\includegraphics[scale=0.6]{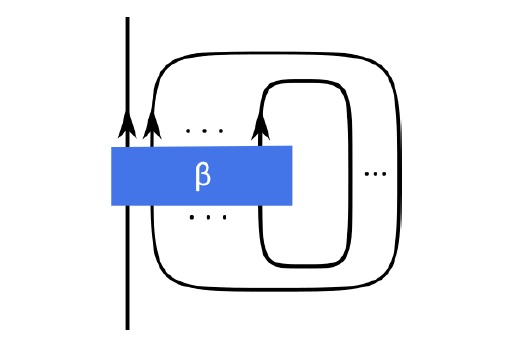}
\caption{Braid $\beta$, more precisely $(1,1)$-tangle setup.}
\end{figure}

\begin{figure}[h!]
\centering
\includegraphics[scale=0.6]{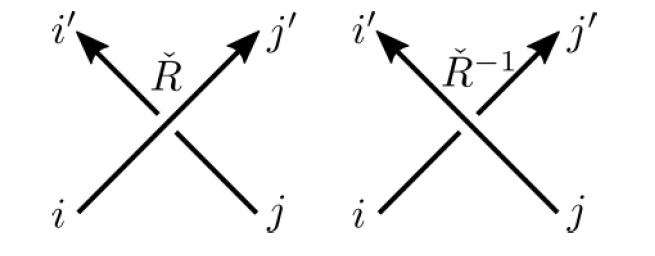}
\caption{$\check{R}$ and $\check{R}^{-1}$ matices for  positive and negative crossings are shown respectively.}
\end{figure}

Inspired by the quantum R-matrix formulation of the colored Jones polynomials, R-matrix formulation for $F_K (x,q)$ was constructed in \cite{P2}. This approach
revealed the the infinite dimensional Verma module structure of quantum group $U_q (sl(2))$ at generic $q$ underlying $F_K (x,q)$. From computational viewpoint, the R-matrix formulation vastly extended classes of knots that $F_K$ can be computed. For example, positive braid knots and positive double twist knots were computed explicitly in \cite{P2}; both of them are infinite families of knots.
\newline
\indent As in the colored Jones polynomials case, the R-matrix approach utilizes a braid presentation $\beta$ of $K$ in Figure 5 and  the R-matrix acts on the infinite dimensional Verma modules of $U_q (sl(2))$ over $\mathcal{F}:= \complex (x^{1/2}, q^{1/2})$. There are two such modules. One of them is the \textit{highest weight Verma module} $V^{h}_{\infty,\lambda}$ with highest weight $\lambda$:
$$
\cdots \rightleftharpoons V^{h}_{\infty} (\lambda -2)  \rightleftharpoons V^{h}_{\infty} (\lambda ),
$$
where the top and the bottom maps are $e$ and $f$ generators of $U_q (sl(2))$, respectively. A basis $\lac v^j \rac_{j\geq 0}$ with $v^j \in V^{h}_{\infty} (\lambda -2j)$ for which the actions of $U_q (sl(2))$ are given by
\begin{align*}
e v^j & = [j]v^{j-1} \\
f v^j & = [\lambda -j]v^{j+1}\\
q^{\frac{h}{2}} v^j & = q^{\frac{\lambda -2j}{2}}v^j,
\end{align*}
where $[k]:=\frac{q^{k/2}-q^{-k/2}}{q^{1/2}-q^{-1/2}}$. The second one is the \textit{lowest weight Verma module} $V^{l}_{\infty,\lambda}$ with lowest weight $\lambda$:
$$
V^{l}_{\infty} (\lambda)  \rightleftharpoons V^{l}_{\infty} (\lambda +2 ) \rightleftharpoons \cdots,
$$
where the top and the bottom maps are $e$ and $f$ generators, respectively. A basis $\lac v_j \rac_{j\geq 0}$ with $v_j \in V^{l}_{\infty} (\lambda +2j)$ for which the action of $U_q (sl(2))$ is given by
\begin{align*}
e v_j & = [-\lambda -j]v_{j+1} \\
f v_j & = [j] v_{j-1}\\
q^{\frac{h}{2}} v_j & = q^{\frac{\lambda +2j}{2}}v_j.
\end{align*}
\begin{Rmk} The color $n=log_q x$ need not be an integer.
\end{Rmk}
\begin{Rmk} The effects of $e$ and $f$ on the bases are interchanged for highest and lowest weight Verma modules.
\end{Rmk}

The quantum R-matrix on the Verma modules for $F_K (x,q)$ for positive and negative crossings (see Figure 6) are given by~\cite{P2},respectively,
\begin{align}
\begin{split}
\check{R}(x)^{i^{\prime},j^{\prime}}_{i,j} & =\delta_{i+j,i^{\prime}+j^{\prime}} q^{\frac{j^{\prime}+j^{\prime}+1}{2}} x^{-\frac{j^{\prime}+j^{\prime}+1}{2}} q^{j j^{\prime} } \begin{bmatrix}
i\\
j^{\prime}
\end{bmatrix}_{q} 
\prod_{1\leq l \leq i-j^{\prime}} \lb 1- q^{j+l}x^{-1} \rb\\
\check{R}^{-1}(x)^{i^{\prime},j^{\prime}}_{i,j} & =\delta_{i+j,i^{\prime}+j^{\prime}} q^{-\frac{i+i^{\prime}+1}{2}} x^{\frac{i+i^{\prime}+1}{2}} q^{-i i^{\prime} } \begin{bmatrix}
j\\
i^{\prime}
\end{bmatrix}_{q^{-1}}
\prod_{1\leq l \leq j-i^{\prime}} \lb 1- q^{-i-l}x \rb\\
\end{split}
\end{align}
These large color R-matrices satisfy the quantum Yang-Baxter equation
$$
\check{R}_{23}\check{R}_{12}\check{R}_{23} = \check{R}_{12} \check{R}_{23}\check{R}_{12}.
$$
\newline
Compactly, the definitions of $F_K$ in terms of a braid closure, and hence the R-matrices is given by
\begin{equation}
F_{K}^{\pm} (x,q) = (x^{1/2}-x^{-1/2})\, Tr_{q}^{\prime}\, \beta_{V^{h,l}_{\infty}},
\end{equation}
where the superscripts $\pm$ denotes positive or negative $x$-expansions of (30) and $ Tr_{q}^{\prime}$ is the reduced quantum trace~\footnote{The reduced refers to opening up a braid as in Figure 5. It is related to the usual quantum trace (see Section 4 of \cite{P2} for detail).}.
\begin{Rmk} An important aspect of the state sum formulation ($F_K =$ Tr) is (absolute) convergence of power series in $x^{\pm 1}$ and $q$. This restricts to classes of knots in which (32) is applicable. For example, positive braid knots, fibered strongly quasi-positive braid knots, and positive double twist knots have been computed.
\end{Rmk}
For the above classes of knots $\beta_K$, we have
$$
F^{-}_K (x,q) = (x^{1/2}-x^{-1/2})\, Tr_{q}^{\prime}\, \beta_{V^{h}_{\infty}}.
$$
In case of mirror knot of $\beta_K$, $V^h$ is replaced by $V^l$ and $F^{-}$ becomes $F^{+}$. The other half can be obtained using Weyl symmetry. 
\begin{Rmk} If two crossing strands are colored by different representations, (31) become functions of $x$ and $y$ variables (see Section 3.2 in \cite{P2} for details).
\end{Rmk}

The above large color R-matrix was generalized to representations of $U_q(sl(N))$ in \cite{Gr}. In the higher rank case, infinite Verma modules form high dimensional lattice, whose dimension depends on the value of $N$. For example, when $N=3$, the dimension of the lattce is two. This is because $U_q(sl(3))$-Verma modules are labeled by $V_{i,j}, i,j\geq 0$. Examples of positive braid knots and homogeneous braid knots have been computed in \cite{Gr}.
\newline

\noindent We move onto link generalization $F_L$ of $F_K (x,q)$.

\subsection{Inverted state sums}

\begin{figure}[h!]
\centering
\includegraphics[scale=0.5]{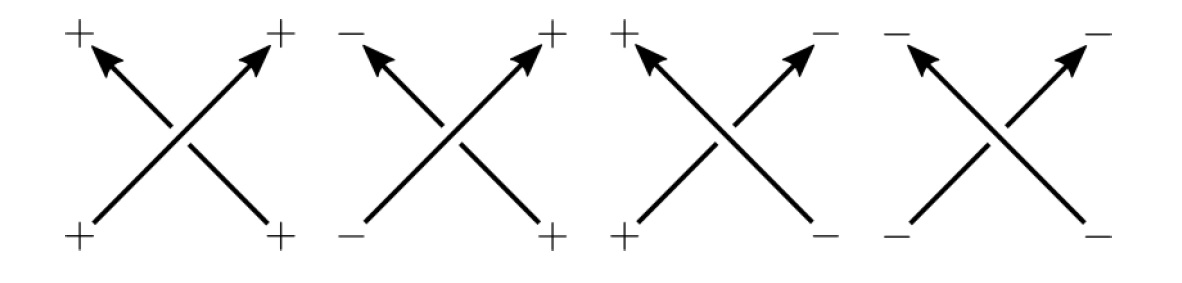}
\caption{Crossing for the inverted state sum.}
\end{figure}

A generalization of (30) to links was achieved in \cite{P3}. Main features of the generalizations are domain extensions of (31) and (32) and the introduction of four building blocks of braid shown in Figure 7. Specifically, the domain of $i,j,i^{\prime},j^{\prime}$ are extended to the set of all integers. In case two strands of a braid are same, the R-matrices are given by
\begin{equation}
\check{R}(x)^{i^{\prime},j^{\prime}}_{i,j} = \begin{cases} 
\delta_{i+j,i^{\prime}+j^{\prime}} q^{\frac{j^{\prime}+j^{\prime}+1}{2}} x^{-\frac{j^{\prime}+j^{\prime}+1}{2}} q^{j j^{\prime} } \begin{bmatrix}
i\\
i - j^{\prime}
\end{bmatrix}_{q} 
\prod_{1\leq l \leq i-j^{\prime}} \lb 1- q^{j+l}x^{-1} \rb , & i\geq j^{\prime} \geq 0\quad or\\
\phantom{1}  & 0 > i \geq j^{\prime}\\
\delta_{i+j,i^{\prime}+j^{\prime}} q^{\frac{j^{\prime}+j^{\prime}+1}{2}} x^{-\frac{j^{\prime}+j^{\prime}+1}{2}} q^{j j^{\prime} } \begin{bmatrix}
i\\
j^{\prime}
\end{bmatrix}_{q} 
\frac{1}{\prod_{0\leq l \leq j^{\prime}-i-1} \lb 1- q^{j-l}x^{-1} \rb }, & j^{\prime} \geq 0 > i\\
0 & otherwise
\end{cases} 
\end{equation}
\begin{equation}
\check{R}^{-1}(x)^{i^{\prime},j^{\prime}}_{i,j}  = \begin{cases} 
\delta_{i+j,i^{\prime}+j^{\prime}} q^{-\frac{i+i^{\prime}+1}{2}} x^{\frac{i+i^{\prime}+1}{2}} q^{-i i^{\prime} } \begin{bmatrix}
j\\
j-i^{\prime}
\end{bmatrix}_{q^{-1}}
\prod_{1\leq l \leq j-i^{\prime}} \lb 1- q^{-i-l} x \rb , & j\geq i^{\prime} \geq 0\quad or\\
\phantom{1}  & 0 > j \geq i^{\prime}\\
\delta_{i+j,i^{\prime}+j^{\prime}} q^{-\frac{i+i^{\prime}+1}{2}} x^{\frac{i+i^{\prime}+1}{2}} q^{-i i^{\prime} } \begin{bmatrix}
j\\
i^{\prime}
\end{bmatrix}_{q^{-1}}
\frac{1}{\prod_{1\leq l \leq i^{\prime}-j-1} \lb 1- q^{-i+l} x \rb}, & i^{\prime} \geq 0 > j\\
0 & otherwise
\end{cases} 
\end{equation}

\noindent In case two strands of a braid are different, we need multicolored R-matrices. The strand associated with $i \rightarrow j^{\prime}$ is assigned $x$ variable whereas the strand associated with $j \rightarrow i^{\prime}$ is assigned $y$ variable. The extended multicolored R-matrices are given by
\begin{equation}
\check{R}(x,y)^{i^{\prime},j^{\prime}}_{i,j} = \begin{cases} 
\delta_{i+j,i^{\prime}+j^{\prime}} q^{\frac{j^{\prime}+j^{\prime}+\frac{1}{2}}{2}} x^{-\frac{i^{\prime}+j^{\prime}+1}{4}} y^{-\frac{3j^{\prime}-i^{\prime}+1}{4}} q^{j j^{\prime} } \begin{bmatrix}
i\\
i - j^{\prime}
\end{bmatrix}_{q} 
\prod_{1\leq l \leq i-j^{\prime}} \lb 1- q^{j+l}y^{-1} \rb , & i\geq j^{\prime} \geq 0\quad or\\
\phantom{1}  & 0 > i \geq j^{\prime}\\
\delta_{i+j,i^{\prime}+j^{\prime}} q^{\frac{j^{\prime}+j^{\prime}+\frac{1}{2}}{2}} x^{-\frac{i^{\prime}+j^{\prime}+1}{4}} y^{-\frac{3j^{\prime}-i+1}{4}}  q^{j j^{\prime} } \begin{bmatrix}
i\\
j^{\prime}
\end{bmatrix}_{q} 
\frac{1}{\prod_{0\leq l \leq j^{\prime}-i-1} \lb 1- q^{j-l}y^{-1} \rb }, & j^{\prime} \geq 0 > i\\
0 & otherwise
\end{cases} 
\end{equation}
\begin{equation}
\check{R}^{-1}(x,y)^{i^{\prime},j^{\prime}}_{i,j}  = \begin{cases} 
\delta_{i+j,i^{\prime}+j^{\prime}} q^{-\frac{i+i^{\prime}+\frac{1}{2}}{2}} x^{\frac{3i^{\prime}-j+1}{4}} y^{\frac{j^{\prime}+i+1}{4}} q^{-i i^{\prime} } \begin{bmatrix}
j\\
j-i^{\prime}
\end{bmatrix}_{q^{-1}}
\prod_{1\leq l \leq j-i^{\prime}} \lb 1- q^{-i-l} x \rb , & j\geq i^{\prime} \geq 0\quad or\\
\phantom{1}  & 0 > j \geq i^{\prime}\\
\delta_{i+j,i^{\prime}+j^{\prime}} q^{-\frac{i+i^{\prime}+\frac{1}{2}}{2}} x^{\frac{3i^{\prime}-j+1}{2}}  y^{\frac{j^{\prime}+i+1}{4}} q^{-i i^{\prime} } \begin{bmatrix}
j\\
i^{\prime}
\end{bmatrix}_{q^{-1}}
\frac{1}{\prod_{1\leq l \leq i^{\prime}-j-1} \lb 1- q^{-i+l} x \rb}, & i^{\prime} \geq 0 > j\\
0 & otherwise
\end{cases} 
\end{equation}
Over and under crossings carry signs, which result in four kinds of crossings as shown in Figure 7. The signs are tied to the highest and lowest weight Verma modules. There are rules for assigning signs to crossings in a braid diagram, more precisely (1,1)-tangle (see Section 1.3 of \cite{P3} for details).
\begin{Rmk} The two variable extended R-matrices (35) and (36) can be obtained from the two variable R-matrices in Section 3.2 of \cite{P2}.
\end{Rmk}

\noindent Using these building blocks, $F_L$ for any homogenous links~\footnote{The definition of homogenous links is that each generator $\sigma_k$ of their braid group appear with either positive or negative powers in a braid word.} can be computed. We state the definition of the inverted state sum.
\begin{Defn} (\cite{P3}) Given a homogeneous braid diagram $\beta$, the \textit{inverted state sum} is
$$
Z^{inv}(\beta) := (-1)^s \sum \lb R\cdots R \rb,
$$
where
$$
s= \sharp (\text{columns with negative crossings}) +  \sharp ( \check{R}^{-1} ).
$$
\end{Defn}

\begin{Thm} (\cite{P3}) For any homogeneous braid link $L$ with a homogeneous braid diagram $\beta_L$, let
$$
F_L := (x^{1/2} - x^{-1/2}) Z^{inv}(\beta_L),
$$
where $x$ is the parameter associated to the open strand. Then $F_L$ is an invariant of $L$. That is, it is independent of the choice of the homogeneous braid representative.
\end{Thm}
The above series $F_L$ is a function of $x_1,\cdots,x_l, q$, where $l$ is the number of components of $L$.
\begin{Rmk} There are more statements in the above theorem. They are written in Section 3.7.
\end{Rmk}

\subsection{Inverted cyclotomic series}

Among several representations of the colored Jones polynomials $J_n (K;q)$ of a knot $K$, there is a particular expression that separates topological and algebraic information of the polynomials and takes values in a completion of the Laurent polynomial ring over the integers~\cite{Hi}. This is often called cyclotomic expansion of the colored Jones polynomials~\footnote{The cyclotomic expansion is also valid for links}. Specifically, a knot colored by n-dimensional irreducible representation $V_n$ of $sl(2)$, its cyclotomic expansion of $J_n (K;q)$ is given by~\footnote{This formula is for 0-framed $K$.}
\begin{align}
\begin{split}
J_n (K;q) & = \sum_{m=0}^{\infty} a_m (K;q) \sigma_m (q^n,q)\\
\sigma_m (q^n,q) & = \prod_{j=1}^m \lb q^n + q^{-n} -q^j - q^{-j} \rb \in Z(U_{\hbar}),
\end{split}
\end{align}
where $a_m (K) \in \intg[q^{\pm 1}] \subset \rational[[\hbar]]$ and $q=e^{\hbar}$. The knot information is completely captured by $a_m (K;q)$ and $\sigma_m $ are basis elements of $Z(U_{\hbar})$. The cyclotomic expansion manifests the integrality property of $J_n (K;q)$. Some examples of $a_m (K;q)$ are
\begin{equation}
a_m (3^{l}_1) = (-1)^m q^{\frac{m(m+3)}{2}}, \quad a_m (3^{r}_1) = (-1)^m q^{-\frac{m(m+3)}{2}},\quad a_m (4_1) = 1.
\end{equation}
\newline
Motivated by (37), an inverted cyclotomic expansion for $F_K (x,q)$ was introduced in \cite{P3}. Two modifications are the domain of $m$ of $a_m (K;q)$ was extended to the set of all integers and $\sigma_m$ was inverted. Specifically, we have the following conjectural formula.
\begin{Conj} (\cite{P3}) For any knot $K$ with $\Delta_K (x) \neq 1$, it has an inverted Habiro series, and it agrees with the $F_K$ in the sense that
\begin{equation}
F_K (x,q) =  -(x^{1/2} - x^{-1/2}) \sum_{m=1}^{\infty} \frac{a_{-m}(K;q)}{\prod_{j=0}^{m-1} \lb x + x^{-1} - q^j - q^{-j} \rb}
\end{equation}
where the right hand side is expanded into a power series in $x$.
\end{Conj}
The extensions of (38) are
$$
a_{-m} (3^{l}_1) = (-1)^m q^{\frac{m(m-3)}{2}}, \quad a_{-m} (3^{r}_1) = (-1)^m q^{-\frac{m(m-3)}{2}},\quad a_{-m} (4_1) = 1.
$$
\begin{Rmk} In case of links, inverted $\sigma_m (x_i,q)$ for each $i$-th link component $L_i$ appears in the denominator of (39).
\end{Rmk}

\noindent The two expressions of $F_K (x,q)$, namely (30) and (39) are related as follows.
\begin{Prop} (\cite{P3}) If we write
\begin{equation*}
-(x^{1/2} - x^{-1/2}) \sum_{m=1}^{\infty} \frac{a_{-m}(K;q)}{\prod_{j=0}^{m-1} \lb x + x^{-1} - q^j - q^{-j} \rb} = x^{1/2} \sum_{j=0}^{\infty} f_j (K;q) x^j,
\end{equation*}
then
\begin{align*}
f_j (K;q) & = \sum_{i=0}^j \begin{bmatrix}
j+i\\
2i
\end{bmatrix} a_{-i-1}(K;q),\\
a_{-i-1}(K;q) & = \sum_{i=0}^j (-1)^{i+j} \begin{bmatrix}
2i\\
i-j
\end{bmatrix} \frac{[2j+1]}{[i+j+1]} f_j (K;q).
\end{align*}
\end{Prop}
We will see in the next section that (39) is useful for predicting a surgery formula for positive integer slopes.


\subsection{Dehn surgery formulas}

Surgery is an indispensable tool in topology. It consists of cutting and gluing manifolds. It provides different perspectives on manifolds and relate them appropriately, thereby revealing multiple presentations of a manifold. This in turn allows to analyze manifolds having sophisticated topology. In three dimensions, Dehn surgery plays an important role. We first review Dehn surgery.
\newline

\indent Let $Y=S^3$ be a closed oriented manifold and $K$ be a knot in $Y$. We carve out a tubular neighborhood of $K$, which is diffeomorphic to $S^1 \times D^2$. This yields in a compact oriented manifold $Y_K$ with a torus boundary. Then glue a solid torus $S^1 \times D^2$ into $Y_K$ along a slope $p/r \in \rational \cup \lac \infty \rac$ via a diffeomorphism. When gluing, a meridian of the solid torus is mapped to $p \mu + r \lambda$ on $\ptl Y_K =T^2$, where $\mu$ is a meridian and $\lambda$ is a longitude of $T^2$. This results in a closed oriented manifold $Y_{p/r}$.
$$
Y_{p/r} = Y_K \cup_{T^2} S^1 \times D^2 .
$$
\begin{Rmk} In case links in $Y$, the above operation is be applied to each component $L_i$ of links with its surgery slope $p_i /r_i$.
\end{Rmk}
There is a classical result regarding closed oriented 3-manifolds and Dehn surgery.
\begin{Thm} (\cite{Li,W}) Any closed connected oriented 3-manifold can be obtained from Dehn surgery on (framed) links in $S^3$. 
\end{Thm} 
In our setting, Dehn surgery establishes a relation between the $F_K$ and $\hat{Z}$. There are multiple (conjectured) surgery formulas covering different regimes of surgery slopes for a knot. The surgery formulas allow to access $\hat{Z}$ of closed oriented 3-manifolds that are beyond plumbed manifolds.
\begin{Thm}(\cite{GM} Plumbed knot surgery) Let $Y$ be the complement of a knot $K$ in an $\hat{Y}=\intg HS^3$, and let $Y_{p/r}$ be the result of Dehn surgery along $K$ with coefficient $p/r\in\rational^{\ast}$. Suppose that both $\hat{Y}$ and $Y_{p/r}$ are negative definite plumbed 3-manifolds. Then the surgery on $K$ yields  
\begin{equation}
\hat{Z}_b (Y_{p/r};q) = \epsilon q^d \mathcal{L}^{(b)}_{p/r} \lsb \lb x^{\frac{1}{2r}} - x^{-\frac{1}{2r}} \rb F_K (x,q) \rsb
\end{equation}
where
$$
\mathcal{L}^{(b)}_{p/r} :  x^{u} q^v  \mapsto
\begin{cases}
q^{-u^2 r/p} q^v , & \text{if}\quad ru-b \in p\intg \\
0,                & \text{otherwise}.
\end{cases}
$$
for some $\epsilon \in \lac \pm 1\rac$ and $d \in \rational$.
\end{Thm}
The above result was conjectured for all knots.
\begin{Conj}(\cite{GM}) Let $K \in S^3$ be a knot, and let $S^3_{p/r}(K)$ be the result of Dehn surgery along $K$ with coefficient $p/r\in\rational^{\ast}$. Then there exists $\epsilon\in \lac \pm 1 \rac$ and $d\in \rational$ such that
\begin{equation}
\hat{Z}_b (S^3_{p/r}(K);q) = \epsilon q^d \mathcal{L}^{(b)}_{p/r} \lsb \lb x^{\frac{1}{2r}} - x^{-\frac{1}{2r}} \rb F_K (x,q) \rsb
\end{equation}
provided that the right hand side of this equation is well defined.
\end{Conj}
The well defined condition is tied to the sign of the surgery slope $p/r$ and the behavior of $f_m (K; q)$ in (30). In case, the right hand side of (41) yields an ill defined result due to positivity of the slope $p/r >0$, an idea of regularization was proposed to obtain a convergent $q$-series in the complex unit disc~\cite{P3}. Specifically, the regularized surgery formulas for positive surgery slopes $+1/r$ and $+p$ are the following.
\begin{Conj}(\cite{P3} Regularized $+\frac{1}{r}$ surgery) When the $+\frac{1}{r}$ surgery formula (41) converges, we need to use (41). When it does not converge, we can regularize it in the following way, as long as the regularization converges:
$$
\hat{Z}_b (S^3_{+\frac{1}{r}}(K);q) = q^{\frac{r+r^{-1}}{4}} \sum_{j=0}^{\infty} f_j (K;q) \lb q^{-r(j+\frac{1}{2}-\frac{1}{2r})^2} -q^{-r(j+\frac{1}{2}+\frac{1}{2r})^2} \rb \lb 1 - \frac{\sum_{|k|\leq j} (-1)^k q^{\frac{k((2r+1)k+1)}{2}}}{f(-q^r, -q^{r+1})}\rb
$$
where
$$
f(a,b):= \sum_{n\in\intg} a^{\frac{n(n+1)}{2}} b^{\frac{n(n-1)}{2}} = (-a;ab)_{\infty}(-b;ab)_{\infty}(ab;ab)_{\infty}
$$
is the Ramanujan theta function.
\end{Conj}
The term in the second parenthesis is the regulator.

\begin{Conj}(\cite{P3} Regularized $+p$ surgery) When the $+p$ surgery formula (41) converges, we need to use (41). When it does not converge, we can regularize it in the following way.
$$
\hat{Z}_b (S^3_{+p}(K);q) = q^{\frac{1}{2}} \sum_{n=0}^{\infty} a_{-n-1}(K) \frac{(-1)^n q^{\frac{n(n+1)}{2}}}{(q^{n+1};q)_n} q^{\frac{b(p-b)}{p}} P^{p,b}_n (q^{-1})
$$
provided that the regularization converges. The polynomials $P^{p,b}_n (q) \in \intg_{\geq 0}[q^{-1}]$ 
\end{Conj}
The above polynomials $P^{p,b}_n (q) \in \intg_{\geq 0}[q^{-1}]$ arise from the following rational function.
$$
\frac{1}{\prod_{j=1}^n \lb x + x^{-1} - q^j - q^{-j} \rb}\bigg|_{x^u \mapsto \delta_{b,u(mod p)}q^{-\frac{u^2}{p}}} = \frac{(-1)^n q^{\frac{n(n+1)}{2}}}{(q^{n+1};q)_n}q^{\frac{b(p-b)}{p}} P^{p,b}_n (q^{-1}). 
$$
A list of $P^{p,b}_n$ is available in \cite{CCKPS}.
\newline

Up to this point, we have been focusing on knot surgeries. In case of links, a formula for Dehn surgery on link along integer surgery slopes has been conjectured.
\begin{Conj}(\cite{P2} Integral link surgery) Let $S^{3}_{p_1,\cdots , p_l} (L)$ be the 3-manifold obtained by $(p_1,\cdots , p_l)$ surgery on $L \subset S^3$, and let $B$ be the $l\times l$ linking matrix defined by
$$
B = \begin{cases}
p_i , \quad \text{if} \quad i=j,\\
lk(i,j), \quad \text{otherwise}
\end{cases}
$$
Let $$
\mathcal{L}^b_{B} : x^u \mapsto   \begin{cases}
q^{-(u,B^{-1}u)}, \quad\text{if}\quad u \in b+ B\intg^l\\
0, \quad\text{otherwise}
\end{cases}
$$
Then 
$$
\hat{Z}_b (S^{3}_{p_1,\cdots , p_l} (L)) = \epsilon q^d \mathcal{L}^b_{B} \lsb \lb x_{1}^{\frac{1}{2}} - x_{1}^{-\frac{1}{2}} \rb \cdots  \lb x_{l}^{\frac{1}{2}} - x_{l}^{-\frac{1}{2}} \rb F_{L}(x_1,\cdots, x_l,q) \rsb
$$
\end{Conj}
for some sign $\epsilon \in \lac \pm 1 \rac$ and $d\in \rational$ , whenever the right hand side makes sense.
\begin{Rmk} A surgery formula for the infinite surgery slope was found~\footnote{This formula would appear in a forthcoming paper by P.Guicardi and M.Jagadale.}.
\end{Rmk}
The above Dehn surgery formulas were generalized in case of presence of the line operators in Section 2.4.
\begin{Defn} (\cite{CCKPS}) Consider the series $F_K (x,q)$ associated to the plumbed knot $K$ with a defect operator along $K$ in the highest weight representation
of $sl(2)$ with highest weight $\nu \vec{\omega}$. We define the corresponding defect invariant for the closed manifold $S_{p/r}(K)$ as
$$
\hat{Z}_b (S_{p/r}(K); W_\nu ;q) = \epsilon q^d  \mathcal{L}^{b+\nu}_{p/r}\lsb ( x^{1/2} - x^{-1/2} )  F_K (x,q) \chi_{\nu} (x^{\frac{1}{2r}}) \rsb
$$
where the sl(2) character $\chi_{\nu}$ (18) and $\epsilon ,d $ are same as in (41). 
\end{Defn}

%

\subsection{Perturbative expansion}

A connection between colored Jones polynomials $J_n (K; q)$ and Alexander polynomials $\Delta (K;x)$ of a knot $K$~\footnote{The knot is 0-framed.} was discovered in \cite{MM,R,R2} and proven in \cite{BG}. This relation appears in the perturbative expansion of the former.
\begin{align*}
J_n (K; e^{\hbar}) & = \frac{1}{\Delta (x)} + \sum_{r=1}^{\infty} \frac{P_r (x)}{\Delta (x)^{2r+1}} \hbar^r\\
                   & = \sum_{m=0}^{\infty} \sum_{j=0}^m c_{m,j}n^j \hbar^m
\end{align*}
where $P_r (x)$ are Laurent polynomials. This expansion is natural from the viewpoint of the Chern-Simons (CS) gauge theory. It is a weak coupling (i.e large CS level $k\in\intg$) regime of the theory. Motivated by the above perturbative expansion, a similar property was conjectured for $F_K$.
\begin{Conj}(\cite{GM}) For a knot $K \subset$ $S^3$, the asymptotic expansion of the knot invariant $F_{K}\big(x,q={\rm e}^{\hbar}\big)$ about $\hbar =0$ coincides with the Melvin--Morton--Rozansky (MMR) expansion of the colored Jones polynomial in the large color limit:
\begin{gather}
\frac{F_{K}\big(x,q={\rm e}^{\hbar}\big)}{x^{1/2}-x^{-1/2}} = \sum_{r=0}^{\infty} \frac{P_{r}(x)}{\Delta_K(x)^{2r+1}}\hbar^r,
\end{gather}
where $x=e^{n\hbar}$ is fixed, n is the color of $K$, $P_{r}(x) \in \mathbb{Q} \big[x^{\pm 1}\big]$, $P_{0}(x)=1$ and $\Delta_K (x)$ is the (symmetrized) Alexander polynomial of~$K$.
\end{Conj}
\noindent A generalization of the above conjecture to links was stated and proved.
\begin{Conj} (\cite{P2}) There is a link invariant $F_L (x_1,\cdots,x_l ,q)$, a series in $x_1,\cdots,x_l$ and $q$ with integer coefficients, where $l$ is the number of components of the link $L$ such that
$$
F_L (x_1,\cdots,x_l ,e^{\hbar}) = \sum_{j=0}^{\infty} \frac{P_j (L;x_1,\cdots,x_l)}{\grd_L (x_1,\cdots,x_l)^{2j+1}} \frac{\hbar^j}{j!}
$$
where the right hand side is the large color expansion of the colored Jones polynomials $J_L (n_1,\cdots, n_l; q=e^{\hbar})$ expanded around $\hbar =0$ while keeping $x_i=q^{n_i} = e^{n_i \hbar}$ fixed for each $1\leq i \leq l$ and $\grd_L $ is the Alexander-Conway function of $L$. 
\end{Conj}
\begin{Thm}(\cite{P3} Theorem 1 (2)) Let $L$ and $F_L$ be as described in Theorem 3.9. Setting $q=e^{\hbar}$, the $\hbar$-expansion of $F_L$ agrees with the MMR expansion of the colored Jones polynomials.
\end{Thm}

The perturbative analysis has been extended to $F_K$ associated with a Lie algebra $sl(3)$ for positive braid knots in \cite{GS}. The perturbation series takes the following form. 
\begin{Thm}(\cite{GS} $F_K$ for $sl(3)$) Let $\beta_K$  be a positive braid knot. Then the reduced quantum trace $\tilde{Tr}^{q}_{V^{1}_x \otimes V^{1}_y} (\beta_K)$ converges in $\intg[q^{\pm 1}][[x^{-1},y^{-1}]]$ and 
$$
F^{sl(3)}_K (x,y,q) = \sum_{i,j\geq 0} f_{i,j}(q) x^{i+\frac{1}{2}}y^{i+\frac{1}{2}}
$$
is well defined knot invariant that satisfies
\begin{equation*}
F^{sl(3)}_K (x,y,e^{\frac{h}{2}}) = \sum_{j=0}^{\infty} \frac{P_j (K;x,y)}{\lb \Delta_K (x) \Delta_K (y) \Delta_K((xy)^{-1}) \rb^{2j+1} } \frac{h^j}{j!} 
\end{equation*}
where $P_j (K;x,y) \in \rational [x^{\pm 1}, y^{\pm 1}]$ and $P_0 =1$.
\end{Thm} 

\begin{Rmk} In \cite{Gr}, knots colored by symmetric representation of $U_q (sl(3))$ was considered. This condition was relaxed in \cite{GS}, which led to the generic R-matrix.
\end{Rmk}

\subsection{Recursion Method}

Another well known property of the colored Jones polynomials of a knot $K$ in $S^3$ is that they are $q$-holonomic~\cite{GL,G2}~\footnote{This property is valid for links as well.}. Specifically, they satisfy a recursion relation
$$
\hat{A}_K (\hat{x}, \hat{y},q) J_n (K;q)= 0,
$$
where $n\in \nat$ is the color of $K$ and $\hat{A}_K $ is called quantum (noncommutative) $A$-polynomial of $K$ and it is a $q$-difference operator of the form
$$
\hat{A}_K (\hat{x}, \hat{y},q) = \sum_{k=0}^d g_k (\hat{x},q) \hat{y}^k.
$$
The operators $\hat{x}$ and $\hat{y}$ acts by
\begin{align*}
\hat{x} J_n (K;q) & = q^n J_n (K;q)\\
\hat{y} J_n (K;q) & = J_{n+1} (K;q)\\
\end{align*}
The above recursion relation enables to find $J_n$ for any color. It was conjectured that $F_K$ also satisfies a recursion relation given by the same $\hat{A}_K$~\cite{GM}. 
\begin{Conj}(\cite{GM}) For any knot $K \subset$ $S^3$, the normalized series $f_{K}(x,q)$ satisfies a linear recursion relation generated by the quantum A-polynomial of $K$ $\hat{A}_K(q,\hat{x},\hat{y})$:
\begin{equation}
\hat{A}_{K}(q, \hat{x},\hat{y}) f_{K}(x,q) = 0,
\end{equation}
where $f_{K}:=F_{K}(x,q)/\big(x^{1/2}-x^{-1/2}\big)$. 
\end{Conj}
\noindent The actions of $\hat{x}$ and $\hat{y}$ are
$$
\hat{x} f_{K}(x,q)= x f_{K}(x,q) \qquad \hat{y}f_{K}(x,q)= f_{K}(xq,q).
$$
\begin{Conj}(\cite{P2}) The link series $F_L (x_1,\cdots, x_l ,q)$ defined in Theorem 3.9 is annihilated by the quantum $A$-ideal annihilating the colored Jones polynomials of $L$.
\end{Conj}
\begin{Thm}(\cite{P3} Theorem 1 (3)) Conjecture 3.28 holds.
\end{Thm}

\subsection{ADO polynomials}

We saw that $\hat{Z}_b (q)$ at roots of unity are related to other topological invariants of 3-manifolds as described in Section 2.6. Similarly, evidence for a connection between $F_K (x,q)$ at roots of unity and ADO polynomials of $K$ was discovered in \cite{GHNPPS}. The latter are colored generalization of the Alexander polynomials and non semisimple quantum invariants~\cite{ADO}. The precise form of the relation is given by the following conjecture.
\begin{Conj} (\cite{GHNPPS}) For any knot $K$ in $S^3$, 
$$
F_K (x, q)|_{q=\zeta_{p}} =  \lb x^{1/2}-x^{-1/2} \rb \frac{ \text{ADO}_{p}(K; x,\zeta_{p})}{\Delta_{K}(x^p)}\qquad \zeta_{p}=e^{i 2 \pi/p},\quad p \in \intg_{+}.
$$ 
\end{Conj}
This conjecture was verified for some values of $p$ for the right-handed trefoil and the figure eight knots~\cite{GHNPPS}. Further evidence for the conjectures including formulas for $ADO_3$ and an algorithm for $ADO_4$ of a family of torus knots was given in \cite{C1}. We record explicit formulas of $\text{ADO}_3$ for $T(2,2s+1)$. They are divided in three types depending on their coefficient pattern.
\begin{enumerate}
	\item For $K=T(2,2s+1)= T(2,3), T(2,9),T(2,15),T(2,21),\cdots$
		\begin{multline*}
	  \text{ADO}_3(x) = \zeta_{3}x^{2s} + \zeta_{3}x^{2s-1} + (\zeta_{3}-\zeta_{3}^{-1})x^{2s-2} - \zeta_{3}^{-1}x^{2s-3} -\zeta_{3}^{-1}x^{2s-4} \\
	        +\zeta_{3}x^{2s-6} + \zeta_{3}x^{2s-7} + (\zeta_{3}-\zeta_{3}^{-1})x^{2s-8} - \zeta_{3}^{-1}x^{2s-9} -\zeta_{3}^{-1}x^{2s-10} \\
					+ \cdots +  (\zeta_{3}-\zeta_{3}^{-1}) + (x \rarw 1/x).
	\end{multline*}	
	
	\item For $K=T(2,2s+1)= T(2,5), T(2,11),T(2,17),T(2,23),\cdots$
		\begin{multline*}
	  \text{ADO}_3(x) = \zeta_{3}^{-1}x^{2s} + \zeta_{3}^{-1} x^{2s-1} + (\zeta_{3}^{-1}-1) x^{2s-2} - x^{2s-3} - x^{2s-4} \\
	        +\zeta_{3}^{-1}x^{2s-6} + \zeta_{3}^{-1} x^{2s-7} + (\zeta_{3}^{-1}-1) x^{2s-8} - x^{2s-9} - x^{2s-10} \\
					+ \cdots -1 + (x \rarw 1/x).
	\end{multline*}	
	
		\item For $K=T(2,2s+1)= T(2,7), T(2,13),T(2,19),T(2,25),\cdots$
		\begin{multline*}
	  \text{ADO}_3(x) = x^{2s} +  x^{2s-1} + (1-\zeta_{3}) x^{2s-2} - \zeta_{3} x^{2s-3} - \zeta_{3} x^{2s-4} \\
	        + x^{2s-6} +  x^{2s-7} + (1-\zeta_{3}) x^{2s-8} - \zeta_{3} x^{2s-9} - \zeta_{3} x^{2s-10} \\
					+ \cdots +1 + (x \rarw 1/x).
	\end{multline*}	
	
\end{enumerate}
All the explicit x terms are polynomials and power of x decreases by two after one cycle of a coefficient combination. Another advancement was an introduction of a refinement of $F_K(x,q)$~\cite{EGGKPS}. It was shown that $F_K(x,q)$ admits two parameter deformations through the superpolynomial~\cite{DGR,FGSS}. This led to a generalization of the above conjecture.
\begin{Conj} (\cite{EGGKPS}) For any knot $K$ in $S^3$, there exists a t-deformation of the symmetric $\text{ADO}_{p}$-polynomial of $K$ for $SU(N)$,
$$
\text{ADO}_{p}^{SU(N)}[K; x,t] : = \lb \Delta_K (x^p, -(-t)^p) \rb^{N-1} \lim_{q \rarw e^{i2\pi /p}}F_K(x,q,a=-q^N/t,t),\qquad p \in \intg_{+}
$$
and $t=-1$ specialization reduces to the original $\text{ADO}_p[K;x]$ (up to rescaling of x). 
\end{Conj}

\noindent From Conjecture 3.31, a refined $\text{ADO}_3$ polynomial for $T(2,2s+1)$, $s \in \intg_{+}$ is
$$
\text{ADO}_{3}[T(2,2s+1);x,t]= (tx)^{2s} + \frac{\zeta_{3}^{-1}}{t}(tx)^{2s-1} + \lb \frac{\zeta_{3}}{t^2} - \zeta_{3}^{-1} \rb (tx)^{2s-2} - \frac{\zeta_{3}}{t}(tx)^{2s-3} -\frac{1}{t^2}(tx)^{2s-4} 
$$
$$
+ (tx)^{2s-6} + \frac{\zeta_{3}^{-1}}{t}(tx)^{2s-7} + \lb \frac{\zeta_{3}}{t^2} - \zeta_{3}^{-1} \rb (tx)^{2s-8} - \frac{\zeta_{3}}{t}(tx)^{2s-9} -\frac{1}{t^2}(tx)^{2s-10} + \cdots + O\lb \frac{1}{tx} \rb,
$$
where $O(1/tx)$-terms are determined by the t-deformed Weyl symmetry of the $\text{ADO}_{p}$ invariant,
$$
\text{ADO}^{SU(2)}_{p} (1/x ,t ) = \text{ADO}^{SU(2)}_{p} ( \zeta_{p}^{-2}t^{-2}x ,t ).
$$
The suppressed polynomial terms follow the same power and coefficient patterns of the previous terms. The three formulas for the original $\text{ADO}_3[T(2,2s+1);x]$ coalesce into one formula by the $t$-deformation.

\subsection{Knot-quiver correspondence}

An interesting connection between the deformed series $F_K (x,a,q)$~\footnote{This series is a series analgoue of the colored HOMFLYPT polynomial.} and the quiver theory~\cite{KRSS,KRSS2} was found in \cite{EGGKPSS}. It was described that $F_K$ can be obtained from so-called \textit{motivic generating series} that characterizes a quiver. We begin by reviewing the quiver side.
\newline

A quiver $Q$ is an oriented graph consisting of a finite set of vertices $Q_0$ and a finite set of arrows between them $Q_1$. (i.e ($Q_0, Q_1$)). An adjacency matrix $C$ of $Q$ is the $m\times m$ matrix with entries $C_{ij}$ equal to the number of arrows from $i$ and $j$, where $m=|Q_0|$. If $C^t = C$, then $Q$ is called a symmetric quiver. A quiver representation is an assignment of a finite dimensional $d_i \in \bm{d}=(d_1,\cdots,d_m)$ is complex vector space to the vertex $i\in Q_0$. and a linear map $\gamma_{ij}: \complex^{d_i} \rightarrow \complex^{d_j}$ to each arrow from vertex $i$ to $j$. A goal in quiver representation theory is to investigate moduli spaces of quiver representations. In case of symmetric quivers, information about the moduli space of representation is encoded in \textit{motivic generating series} defined as
\begin{equation}
P_Q (\bm{x},q) : =\sum_{d_1,\cdots, d_m \geq 0}(-q^{1/2})^{\bm{d}.C.\bm{d}} \prod_{i=1}^m \frac{x_{i}^{d_i}}{(q;q)_{d_i}},
\end{equation}
where
$$
(z;q)_{n} = \prod_{k=0}^{n-1} (1-zq^k).
$$
In \cite{Ku}, it was shown that the knot-quiver correspondence can be generalized to knot complements of torus knots $T(2,2n+1)$. Specifically, this involves data of a symmetric $Q$, integers $n_i$, and half-integers $a_i ,l_i, i\in Q_0$ to a knot complement $M_K = S^3 \backslash \nu(K)$. Then the deformed $F_K$ can be obtain from (44) by
\begin{equation}
F_K(x,a,q) = P_Q (\boldsymbol{x},q)\bigg|_{x_i = x^{n_i}a^{a_i}q^{l_i}} = \sum_{\bm{d}\geq 0} (-q^{1/2})^{\bm{d}.C.\bm{d}} \frac{x^{\bm{n}.\bm{d}}a^{\bm{a}.\bm{d}}q^{\bm{l}.\bm{d}}}{(q)_{\bm{d}}}.
\end{equation}
where $(q)_s = (q;q)_s$.
\begin{Rmk} We note that the deformed $F_K (x,a,q)$ is associated with an abelian branch of A-polynomial of $K$. For $F_K$ associated with other branches, see Section 3 of \cite{EGGKPSS}.
\end{Rmk}
\noindent The general quiver form of $F_{T(2,2p+1)} (x,a,q)$ is given by~\cite{EGGKPS}
$$
F_{T(2,2p+1)}(x,a,q) = \sum_{d_1,\cdots,d_{2p+2}\geq 0} (-q^{1/2})^{\bm{d}.C.\bm{d}} \frac{x^{\bm{n}.\bm{d}} a^{\bm{a}.\bm{d}} q^{\bm{q}.\bm{d}-\frac{1}{2}\sum_i C_{ii}d_i} }{(q)_{\bm{d}}}
$$
$$
C = \begin{pmatrix}
\bm{1}_{2p} - \bm{D}  & \bm{-1} & \bm{0}\\
\bm{-1} & 1 & 0\\
\bm{0} & 0  & 0\\
\end{pmatrix}
$$
\begin{align*}
\bm{n} & = \lb 1,1,3,3,\cdots, 2p-1, 2p-1,1,1\rb\\
\bm{a} & = \lb 1,0,\cdots,1,0,0,0 \rb\\
\bm{q} & = \lb 0,1,\cdots,0,1,1,1 \rb = 1- \bm{a},
\end{align*}
where $\bm{−1}, \bm{0}$ denote constant vectors of appropriate size, $\bm{1}_{2p}$ is the identity matrix and $\bm{D}_{ij}=min(i,j)-1$ for $1\leq i,j \leq 2p$.
\newline

The converse direction, namely extracting a quiver structure from $F_K(x,a,q)$ was shown in \cite{EGGKPSS}. For example, we start	 from $F_K(x,a,q)$ the left handed trefoil in the following form.
$$
F_{3_1} (x,a,q) = \sum_{k\geq 0} x^k q^k \frac{(x;q^{-1})_k (aq^{-1};q)}{(q)_k}.
$$
We next apply the following identity to $(aq^{-1};q)$~\footnote{It is Lemma 4.5 in \cite{KRSS2}.}.
\begin{align*}
\frac{(x)_{d_1 +\cdots + d_k}}{(q)_{d_1}\cdots (q)_{d_k}} & = \sum_{\alpha_1 + \beta_1 = d_1} \cdots  \sum_{\alpha_k + \beta_k = d_k} \frac{1}{(q)_{\alpha_1}\cdots (q)_{\alpha_k}(q)_{\beta_1}\cdots (q)_{\beta_k}} \times\\
& \times (-x)^{\alpha_1 +\cdots + \alpha_k } q^{\frac{1}{2}\lb \alpha_{1}^2 + \cdots + \alpha_{k}^2 \rb }q^{-\frac{1}{2}\lb \alpha_{1} + \cdots + \alpha_{k} \rb } q^{\sum_{i=1}^{k-1} \alpha_{i+1}(d_1 + \cdots + d_k )}
\end{align*}
Then we get
\begin{equation}
F_{3_1} (x,a,q) = \sum_{d_{1}^{\prime}, d_{2}^{\prime} \geq 0} (-1)^{d_{1}^{\prime}}  x^{d_{1}^{\prime} + d_{2}^{\prime}} a^{d_{1}^{\prime}} q^{d_{2}^{\prime}} q^{(d_{1}^{\prime\, 2}- d_{1}^{\prime})/2} \frac{(x;q^{-1})_{d_{1}^{\prime}+d_{2}^{\prime}}}{(q)_{d_{1}^{\prime}} (q)_{d_{2}^{\prime}}}
\end{equation}
After using the following identity,
\begin{equation}
(x;q^{-1})_d = (xq^{1-d};q)_d = \frac{(xq^{1-d};q)_{\infty}}{(xq;q)_{\infty}} =\sum_{i,j} (-1)^i x^{i+j}q^{(i^2 -i)/2} q^{i+j} q^{-di} \frac{1}{(q)_i (q)_j}.
\end{equation}
we arrive at a quiver form
$$
F_{3_1} (x,a,q) = \sum_{d_1,\cdots,d_4 \geq 0} (-q^{1/2})^{\sum_{i,j=1}^4 C_{ij}d_i d_j} \prod_{i=1}^4 \frac{x_{i}^{d_i}}{(q)_{d_i}},
$$
\begin{equation}
C = \begin{pmatrix}
0 & 0 & 0 & -1\\
0 & 1 & 0 & -1\\
0 & 0 & 0 & 0\\
-1 & -1 & 0 & 1\\
\end{pmatrix},\qquad
\begin{pmatrix}
x_1 \\
x_2 \\
x_3 \\
x_4 \\
\end{pmatrix} = \begin{pmatrix}
xq \\
xaq^{-1/2} \\
xq \\
xq^{1/2}\\
\end{pmatrix}.
\end{equation}
An alternative identity that can be applied to (46) is
\begin{align}
\begin{split}
\frac{(x;q^{-1})_{d_1 + \cdots + d_n}}{\prod_{i=n}^1 (q;q)_{d_i}} & = \sum_{\alpha_1 + \beta_1 = d_1 }\cdots \sum_{\alpha_n + \beta_n = d_n } (-q^{1/2})^{\beta_{1}^2 + \cdots + \beta_{n}^2 + 2\sum_{i=1}^{n-1} \beta_{i+1} (d_1 + \cdots d_n )}\\
& \times \frac{\lb x q^{1/2 - \sum_i \alpha_i - \sum_i \beta_i} \rb^{\beta_1 + \cdots + \beta_n}}{(q)_{\alpha_1} (q)_{\beta_1}\cdots (q)_{\alpha_n} (q)_{\beta_n} }.
\end{split}
\end{align}
An application of this identity yields
\begin{equation}
C = \begin{pmatrix}
0 & 0 & -1 & -1\\
0 & 1 & 0 & 0\\
-1 & 0 & -1 & -1\\
-1 & 0 & -1 & 0\\
\end{pmatrix},\qquad
\begin{pmatrix}
x_1 \\
x_2 \\
x_3 \\
x_4 \\
\end{pmatrix} = \begin{pmatrix}
xq \\
xaq^{-1/2} \\
x^2 q^{3/2} \\
x^2 a\\
\end{pmatrix}.
\end{equation}
\begin{Rmk} The final forms (48) and (50) are connected by operations on vertex of quivers (see \cite{EKL} for details.)
\end{Rmk}
A closely related result in \cite{EGGKPSS} is that colored HOMFLYPT polynomials $P_r(K;a,q)$ of knots $K$ colored by symmetric representation $S^r$ in a quiver form~\footnote{The subscript $r$ in $P_r$ refers to $S^r$. There is a conjecture regarding expressing the generating function of the colored HOMFLYPT polynomials in a quiver form in \cite{KRSS2}.} can be used to obtain $F_K(x,a,q)$. This requires a framing $K$ in a particular way. Let $Q$ be a quiver corresponding to $K$ and $\bm{a}$ and $\bm{l}$ be the vectors. Suppose that
$$
-C_{\text{min}} \leq C_{ij} \leq C_{max},\qquad i,j=1,\cdots, m,
$$
where $C_{\text{min}}, C_{max} \geq 0$. Next, permute rows and columns of $C$ such that $C_{11}=C_{\text{min}}$ and $C_{mm}=C_{max}$. Expressing $P_r(K;a,q)$ in a quiver form as
\begin{equation}
P_r(K;a,q) = \sum_{d_1 + \cdots + d_k = r} (-1)^{\sum C_{ii}d_i } a^{\bm{a}.\bm{d}}  q^{\bm{l}.\bm{d}} q^{\frac{1}{2} \bm{d}.C.\bm{d}} \frac{(q)_r}{\prod_{i=1}^k (q)_{d_i}}.
\end{equation}
To convert (51) to $F_K(x,a,q)$, we framed $K$ by $C_{\text{min}}$, which amounts to multiplying $P_r$ by $q^{C_{min}(r^2-r)/2}$ and set $q^r =x$.
\begin{align}
\begin{split}
F_{K^{f=C_{\text{min}}}}(x,a,q) & = (-1)^{r C_{\text{min}}} a^{r a_1}q^{r l_1} \sum_{d_1,\cdots, d_m} (-1)^{\sum_{i\geq 2} (C_{ii}+C_{\text{min}} ) d_i} a^{\sum_{i\geq 2} (a_i -a_1)d_i}\\
& \times q^{\sum_{i\geq 2} (l_i - l_1) d_i} x^{\sum_{i\geq 2} (C_{1i} + C_{\text{min}} ) d_i} q^{\frac{1}{2}\sum_{i,j\geq 2} (C_{ij}-C_{i1}-C_{1j}+C_{11})d_i d_j}\\
& \times \frac{(x;q^{-1})_{d_2 + \cdots + d_k}}{\prod_{i=2}^k (q)_{d_i}}.
\end{split}
\end{align}
The last step is applying (47) or (49) to (52).
\begin{Rmk} It was shown that $F_K$ in terms R-matrices and inverted cyclotomic series in the previous sections can be transformed into the quiver form as well~\cite{EGGKPSS}. 
\end{Rmk}

\subsection{Examples}

A variety of examples of $F_K$ have been computed. We summarize a subset of them.
\begin{Thm} (\cite{GM}) Let $s, t > 1$ with gcd$(s; t) = 1$. For the positive torus knot $K = T(s, t)$, the series $F_K (x,q)$ is given by 
$$
F_K (x,q) = q^{\frac{(s-1)(t-1)}{2}} \frac{1}{2} \sum_{\substack {m =1 \\ \text{odd} } }^{\infty} \epsilon_m  q^{\frac{m^2 - (st-s-t)^2}{4st}}  \lb x^{m/2} - x^{-m/2} \rb
$$
where
$$
\epsilon_m =
\begin{cases}
+1 , & \text{if}\quad m \equiv st + s+t \quad\text{or}\quad st -s-t\quad  mod\, 2st\\
-1,  &  \text{if}\quad m \equiv st + s-t \quad\text{or}\quad st -s+t\quad  mod\, 2st\\
0,                 & \text{otherwise}.
\end{cases}
$$
\end{Thm}
The above example $F_{T(s, t)} (x,q)$ is monomials in $q$. It is the only knot of that feature to the best of the author's knowledge.\\
In case of mirror torus knots $T(s,-t)$,
$$
F_{T(s,-t)} (x,q) = F_{T(s,t)} (x,q^{-1}).
$$
\noindent\underline{Figure eight $4_1$} This was first hyperbolic knot computed via the recursion method in Section 3.7~\cite{GM}. A closed form formula was obtained using the R-matrix in Section 3.2 in \cite{P3}. Some of $f_m (4_1;q)$ are
\begin{align*}
f_1 & = 1\\
f_3 & = 2\\
f_5 & = q + 3+1/q\\
f_7 & = 2q^2 + 2q + 5 + 2/q + 2/q^2\\
f_9 & = q^4 + 3q^3 + 4q^2 + 5q + 8 + 5/q + 4/q^2 + 3/q^3 + 1/q^4
\end{align*}
We observe that $f_m (1/q)=f_m (q)$ reflecting amphichirality property of the knot.
\newline

\noindent\underline{Positive double twist knots}~\cite{P2} $K=K_{m,p},\, m,p>0$ full twists,
\begin{align*}
F_{K_{m,p}}^{+} & = \lb x^{1/2}-x^{-1/2} \rb q^{-1}x \sum_{0\leq n_1 \leq \cdots \leq n_{2mp-1}} (xq^{-1};q)_{n_{2mp-1}} (-1)^{n_{2mp-1}} q^{\binom{n_{2mp-1}+1}{2}}\\
                & \times \prod_{\substack{1\leq i < j\leq 2mp-1\\ m \nmid i}} q^{-\epsilon_{i,j,m}n_i n_j} \prod_{i}^{2p-1} (-1)^{n_{mi}} x^{(-1)^{i+1}n_{mi}} q^{-\binom{n_{mi}+1}{2}} \prod_{i=1}^{2mp-2} q^{n_i n_{i+1} - \gamma_{i,m}n_i} \begin{bmatrix}
		n_{i+1}\\
		n_i
		\end{bmatrix}
\end{align*}
where $\epsilon_{i,j,m}$ and $\gamma_{i,m}$ are sign functions. 
\begin{Rmk} The above family of knots include left handed trefoil ($K_{1,1}$) and $5_2 (K_{2,1}$). The series $F_K$ of the latter has $\lac f_m (5_2;q) \rac$ as Laurent power series $\intg [q^{-1},q]]$.
\end{Rmk} 
\begin{Rmk} There is also a formula for $K_{m+\frac{1}{2},p}$ family (see Section 4.4.1 in \cite{P2} for details).
\end{Rmk} 
\noindent\underline{Cable knots}~\cite{C3,C4} Combining the torus knots and the figure eight knot from the above examples, infinite families of cable knots (a class of satellite knots) were analyzed using the recursion method in Section 3.7. Specifically, $F_K$ of
$$
K= C_{(2,2w+1)} (4_1), \qquad  C_{(3,3w+1)} (4_1), \qquad |w|>3,
$$
were computed. Their coefficient functions $f_m (K;q)$ are linear combinations of coefficient functions $h_m (4_1;q)$ of $F_{4_1}$~\footnote{A cabling formula for $C_{(ns,nt)}(T(b,c))$ was found and would appear in a forthcoming paper by P.Guicardi and M.Jagadale.}.

\section{An extension to Lie superalgebras}

\subsection{The super series}

We review the $q$ power series invariant of closed oriented 3-manifolds associated with a Lie superalgebra $sl(2|1)$ introduced in \cite{FP}.
\newline

A non semi-simple quantum invariant of closed oriented 3-manifolds $Y$ associated with $U^{(H)}_q (sl(2|1))$ at a root of unity of odd order was constructed in \cite{H}. Core ingredients of the construction are a non semi-simple ribbon category of simple finite dimensional representations of $U^{(H)}_q (sl(2|1))$ from \cite{CGP} and the modified quantum dimension. The data for the quantum invariant of $Y$ are the root of unity of odd order $q = e^{i4\pi /l }$, odd $l\geq 3$ and a 1-cocycle, 
$$
\omega \in H^{1} (Y ; \complex / \intg \times \complex / \intg) \backslash \bigcup_{i=1}^3 H^{1} (Y ; C_i),
$$
\begin{align*}
C_1 & = \lac (X,Y) \in \complex / \intg \times \complex / \intg | 2X = 0\, \text{mod}\, 1 \rac\\
C_2 & = \lac (X,Y) \in \complex / \intg \times \complex / \intg | 2Y = 0\, \text{mod}\, 1 \rac\\
C_3 & = \lac (X,Y) \in \complex / \intg \times \complex / \intg | 2(X+Y) = 0\, \text{mod}\, 1 \rac.
\end{align*}
Then the non semi-simple quantum invariant is denoted by
\begin{equation}
N_{l} (Y, \omega ) \in \complex .
\end{equation}
\noindent In case of a particular class of 3-manifolds called plumbed manifolds $Y_{\Gamma}$, it was shown in \cite{FP} that (53) decomposes into $q$-power series:
\begin{equation}
\hat{Z}^{sl(2|1)}_{b,c}[ Y_{\Gamma}; q]  \in \rational + q^{\Delta_{b,c}}\intg[[q]],\qquad |q|<1,
\end{equation}
$$
(b,c) \in H_1 (Y ; \intg) \times H_1 (Y;\intg) \cong Spin^c (Y) \times Spin^c (Y),
$$
where $\Delta_{b,c} \in \rational$ and $Spin^c (Y)$ is $Spin^c$ structures on $Y$~\footnote{Its definition is a lift of the structure group $SO(3)$ of the tangent bundle $TY$ of $Y$ to $Spin^c (Y)$ group.}. This $q$ series is an analytic continuation of (53) into the complex unit disk. The decomposition of (53) is given by
\begin{multline}
N_{l} (Y(\Gamma), \omega ) = \frac{\prod\limits_{i\in V} \lb e^{i2\pi \mu^{i}_1} - e^{-i2\pi \mu^{i}_1} \rb^{\text{deg(i)-2}}}{l |Det B|}\times\\\\
\times \sum_{\substack{ \beta, \gamma \in \intg^L /B\intg^L \\ b, c \in B^{-1}\intg^L / \intg^L }} e^{i2\pi l \gamma^t B^{-1} \beta + i4\pi (b- \mu_2 )^t \gamma + i2\pi(c- (\mu_1 + \mu_2 ))^t \beta} (-1)^{\Pi} \hat{Z}^{sl(2|1)}_{b,c}[Y(\Gamma); q]\bigg|_{q \rightarrow \zeta^2},
\end{multline}
where $\zeta = q^{1/2}$, and $(\mu^{i}_1 , \mu^{i}_2) \in \rational / \intg \times  \rational / \intg$. Furthermore, 
\begin{equation}
\hat{Z}^{sl(2|1)}_{b,c}[Y(\Gamma); q] = (-1)^{\pi}  \prod_{v \in V} \int\limits_{\Omega} \frac{dy_v}{i2\pi y_v}\frac{dz_v}{i2\pi z_v} \lb \frac{y_v - z_v}{(1-y_v)(1-z_v)} \rb^{2- \text{deg}(v_s)} \bigg|_{\alpha_i} \Theta_{b,c}(\vec{y},\vec{z},q),
\end{equation}
\begin{equation*}
\Theta_{b,c}  = \sum_{\substack{ \vec{l_1} \in B \intg^s + \vec{b} \\ \vec{l_2} \in B \intg^s + \vec{c}}} q^{(\vec{l_1}, B^{-1} \vec{l_2}) } \prod_{v \in V} y_{v}^{l_{1,v}}z_{v}^{l_{2,v}}, 
\end{equation*}
where $V$ is the vertex set of $\Gamma$, $\pi$ is the number of positive eigenvalues of $B$ and $\alpha_i$ indicates a choice of chamber. And $\Omega$ is an integration contour. Moreover, the variables are $y_I := e^{(e_1 -f_1)(h_I)}$ and $z_I := e^{(e_1 -f_2)(h_I)}$ and are coordinates on the maximal torus, where $e_i$ and $f_1$ are roots. In contrast to $\hat{Z}_b $ associated with the classical Lie algebras~\cite{GPPV, P1}, the super $\hat{Z}_{b,c}$ (56) carries two labels $(b,c)$.
\begin{Rmk} The above integrations are equivalent to picking constant terms in the variables.
\end{Rmk}
\noindent \underline{Generic plumbing graphs} A notion of genericity of plumbing graphs was introduced in \cite{FP}. The definition states that, for a plumbing graph containing at least one vertex whose degree is larger than two, the graph does not admit splitting $V|_{\text{deg}\neq 2} = U \sqcup W$ such that if $i\in U$ and $j\in W$, then $B^{-1}_{ij}=0$, where $V|_{\text{deg}\neq 2}$ is the set of vertices whose degrees are not two.
\newline

\noindent \underline{Good Chambers} The integration contour $\Omega$ in (56) is equivalent to a choice of an expansion chamber $\alpha_i$. In order for (56) to yields a well defined power series, a (generic) plumbing graph containing at least one vertex of degree larger than two must have good chambers. The existence condition of good chambers is given in \cite{FP}: If there exists a vector
$$
\alpha_i = \pm 1, \qquad i \in V|_{\text{deg}\neq 2}
$$
such that 
$$
X_{ij} : = -B^{-1}_{ij} \alpha_i \alpha_j,\quad i,j \in  V|_{\text{deg}> 2}
$$
is \textit{copositive} and
\begin{align}
B^{-1}_{ij} \alpha_i \alpha_j & \leq 0 ,\qquad \forall i \in  V|_{\text{deg}=1},\qquad j \in  V|_{\text{deg}\neq 2}\\
B^{-1}_{ij} \alpha_i \alpha_j & < 0 ,\qquad \forall i,j \in  V|_{\text{deg}=1},\qquad i \neq j
\end{align}
The matrix $X$ is \textit{copositive} if for any vector $v$ such that $v_i \geq 0, \forall i$, with at least one $v_i \neq 0$ and have $\sum_{i,j} X_{ij} v_i v_j >0$. 
\newline

If a good chamber $\alpha$ exists for a generic plumbing graph, then there are two of them and the domains of $y_i$ and $z_i$ corresponding to a vertex $v_i$ are given by
$$
deg(i) = 1 : \begin{cases}
|y_i|^{\alpha_i} < 1\\
|z_i|^{\alpha_i} > 1
\end{cases} \qquad
deg(i) > 2 : \bigg|\frac{y_i}{z_i}\bigg|^{\alpha_i} < 1.
$$
This translates to the following allowed expansions. For vertices $i \in V$ of degree $\text{deg}(i) = 2+ K > 2$,  expansions are
\begin{equation}
\lb \frac{(1-y_i)(1-z_i)}{y_i - z_i} \rb^K = 
\begin{cases}
(z_i -1)^K (1- y_{K}^{-1})^K \sum\limits_{r = 0}^{\infty}  \frac{(r +1)(r +2)\cdots (r +K-1)}{(K-1)!} \lb\frac{z_i}{y_i}\rb^r, & |y_i | > |z_i|\\
(1- z_{K}^{-1})^K (1- y_{K})^K \sum\limits_{r = 0}^{\infty}  \frac{(r +1)(r +2)\cdots (r +K-1)}{(K-1)!} \lb\frac{y_i}{ z_i }\rb^r, & |z_i | > |y_i|.\\
\end{cases}
\end{equation}
For vertices $i \in V$ of degree $\text{deg}(i)=1$, expansions are
\begin{equation}
\frac{y_i - z_i}{(1-y_i)(1-z_i)} = 
\begin{cases}
1+ \sum\limits_{r=1}^{\infty} y_{i}^{r} + \sum\limits_{r=1}^{\infty} z_{i}^{-r} , & |y_i | < 1 , |z_i|>1 \\
-1- \sum\limits_{r=1}^{\infty} y_{i}^{-r} - \sum\limits_{r=1}^{\infty} z_{i}^{r} , & |y_i| > 1 , |z_i|<1 . \\
\end{cases}
\end{equation}
Several remarks are in order.
\begin{Rmk} Other domains of expansions are $|y_i|,|z_i| >1$ and $|y_i|,|z_i| < 1$. However, they are incompatible with good chambers~\cite{FP}.
\end {Rmk}
\begin{Rmk} Generic property of plumbing graphs ensures that (56) is independent of a choice of good chamber $\alpha_i$.
\end {Rmk}
\begin{Rmk} In (54), $\rational$ constant comes from regularizing a diverging constant. We will see in the origin of the diverging constant in Section 5 and 6.
\end {Rmk}

\noindent For general closed oriented 3-manifolds $Y$, we have the following conjecture.
\begin{Conj} The quantum invariant $N_{l} (Y, \omega )$ (53) of closed oriented 3-manifolds that are rational homology spheres ($b_1 (Y) =0$) admits a decomposition in terms of the super $\hat{Z}_{b,c}$. 
$$
N_{l} (Y, \omega ) = \frac{\pm T(2\lsb \omega \rsb)}{l|H_1 (Y;\intg)|} \sum_{\substack{ \beta, \gamma \in H_1 (Y;\intg) \\ b, c \in H^1 (Y;\rational/\intg ) }} e^{i2\pi l \text{lk}(\beta,\gamma) + i4\pi(b-\omega_2)(\gamma) + i2\pi (c-(\omega_1 + \omega_2))(\beta)} \hat{Z}_{b,c}(Y;q)\bigg|_{q \rightarrow \zeta^2}
$$
where $T$ is the Reidemeister torsion of the $U(1)$ flat connection $\lsb 2\omega_1 \rsb := (2\omega_1 \, \text{mod}\, H^1 (Y;\intg)) \in H^1 (Y;\rational/\intg)$.
\end{Conj}

\begin{Rmk} The above $\pm$ reflects the sign ambiguity in the definition of the torsion.
\end{Rmk}

\begin{Prop} (\cite{C5}) The super $\hat{Z}_{b,c} (q)$ (56) is invariant under the Kirby-Neumann moves in Figure 1.
\end{Prop}

\noindent\underline{Examples} We list a few examples~\cite{FP}.
\begin{align*}
\hat{Z}(S^3;q) & = -\frac{1}{6} +2 \sum_{m=1}^{\infty} d(m) q^m\\
               & = -\frac{1}{6} +2 \lb q + 2q^2 + 2q^3 +3q^4 + 2q^5 + 4q^6 + \cdots \rb
\end{align*}
where $d(m)$ is the number of divisors of $m$.
$$
\hat{Z}_{b,c}(L(p,1);q)  = c_{b,c} + 2 q^{\frac{(p-b)(p-c)}{p}-(p-b)}\sum_{k=1}^{\infty} \frac{q^{ck}}{1-q^{pk-(p-b)}}
$$
$$
c_{b,c} = \begin{cases}
1 + 2p \zeta(-1) + 2 \zeta(0),  &  b=c=0              \quad \text{mod}\, p\\
p \zeta(-1,b/p) + \zeta(0,b/p), & b\neq 0, c=0      \quad \text{mod}\, p\\
p \zeta(-1,c/p) + \zeta(0,c/p), & b=0,  c\neq 0, \quad \text{mod}\, p\\
0,                              & b,c\neq 0 \quad\text{mod}\, p
\end{cases}
$$
where $1 \leq b,c \leq p$ and $\zeta(s,x) := \sum_{n\geq 0 } 1/(n+x)^s$ is the Hurwitz zeta function.
\newline

For general $gl(N|M)$, we write the formula manifests the algebraic structure of the superalgebra. Specifically, its super $\hat{Z}_{a}$ for a plumbed manifold with its adjacency/linking matrix $B$ is~\cite{FP}
$$
\hat{Z}_{a}  =  (-1)^{|\Delta_+ | \pi} q^{\frac{3\sigma - Tr B}{2} (\rho,\rho)} \int_{\Omega} \prod_{I \in V} dh_I  D_{g}(\alpha,h_I)^{2-\text{deg}(I)}\sum_{\vec{n} \in B\intg^L \otimes \Lambda + \frac{a}{2}} q^{-\frac{1}{2} \vec{n}^t (B^{-1} \otimes K) \vec{n}}  e^{ \vec{n} (\oplus_{I} h_I )}
$$
$$
D_g (\alpha,h_I) = \prod_{\alpha \in\Delta_+ } \lb e^{\alpha(h_I)/2} - e^{-\alpha(h_I)/2} \rb^{\epsilon(\alpha)},
$$
where $\epsilon(\alpha)= \pm 1$ for even/odd roots $\alpha$~\footnote{$D_{g}$ appears in the Weyl super character formula.}, $dh_I$ is the normalized measure on a maximal torus of $gl(N|M)$, $\Lambda$ is the root lattice, $K:\Lambda \otimes \Lambda \rightarrow \intg$ is the Killing form, $\Delta_{+}$ is the set of positive roots, $\rho$ is the Weyl vector, and $\pi$ is the number of positive eigenvalues of $B$.

\subsection{Supergroup Chern-Simons theory}

\begin{figure}
\centering

  \includegraphics[scale=1.2]{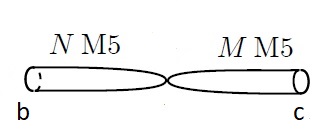}
	\qquad
%
  \includegraphics[scale=0.7]{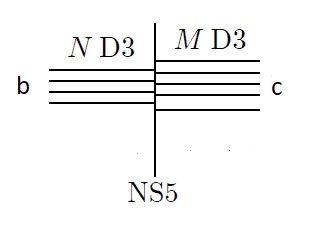}
\caption{The cigars of the Taub-NUT space of 11-dimensional spacetime that are wrapped by the branes (left). The brane system of the Type IIB theory (right). The labels $b$ and $c$ are the asymptotic boundary conditions taking values in  $H_1 (M^3 ;\intg)^N \times H_1 (M^3 ;\intg)^M$ for $U(N|M)$.}
\end{figure}

We realize the Chern-Simons theory on $Y$ associated with a Lie supergroup $U(N|M)$ as a worldvolume theory in string/M-theory~\cite{FP, MW} (see also \cite{V}). This allows us to predict an explicit formula of super $\hat{Z}^{U(N|M)}_{b,c}$ of $Y$.
\newline

We begin with a brane system in a 11-dimensional spacetime (ST) in M-theory. We take the 10d spatial geometry to be a cotangent bundle of a 3-manifold $M^3=Y$ and the 6-dimensional Taub-NUT (TN) space. The former is assumed to be a rational homology sphere. The latter looks like two cigars whose tips are joined at an origin. Away from the tip, the geometry looks like $S^{1}_M \times \real^3$, where the circle is taken to be the M-theory circle. Near tip geometry looks like $\complex^2 \cong \real^4$.
\begin{center}
\begin{tabular}{c c c c c c c c}
11D ST      &   $S^{1}_{\text{t}}$ &  $\times$ &  $T^{\ast}M^3$ & $\times$ & Taub           & $-$       & NUT\\ 
M M5 branes & $S^{1}_{\text{t}}$   & $\times$  & $M^3$          & $\times$ & $\complex$     & $\times $ & $\lac 0 \rac$\\  
N M5 branes & $S^{1}_{\text{t}}$   & $\times$  & $M^3$          & $\times$ & $\lac 0 \rac$  & $\times$  &  $\complex$
\end{tabular}
\end{center}
where $S^{1}_{\text{t}}$ is a time circle. The two stacks of M5 branes wrap the indicated parts of the spacetime as shown in Figure 8. The copies of $\complex $ are part of the TN space. This spacetime geometry has symmetries from the TN space, $U(1)_q \times U(1)_R$~\footnote{If $M^3$ has a circle fiber, for example, a Seifert fibered manifold, then an extra symmetry group $U(1)$ exists.}. We next shrink $S^{1}_M$ to reduce to 10 dimensional spacetime. This process lands us in type IIA string theory and the brane system becomes
\begin{center}
\begin{tabular}{c c c c c c}
Type IIA 10D ST  & $S^{1}_{\text{t}}$ &  $\times $ & $T^{\ast}M^3 $ & $\times $ & $\real^3$\\
1 D6 brane       & $S^{1}_{\text{t}}$ &  $\times $ & $T^{\ast}M^3 $ & $\times $ & $\lac 0 \rac$\\
M D4 branes      & $S^{1}_{\text{t}}$ &  $\times $ & $M^3$          & $\times $ & $\real_{+}$\\  
N D4 branes      & $S^{1}_{\text{t}}$ &  $\times $ & $M^3$          & $\times $ & $\real_{-}$ 
\end{tabular}
\end{center}
The M5 branes are transformed into the D4 branes. The D6 brane appears as a consequence of the Taub-NUT space. We apply T-duality along $S^{1}_{\text{t}}$ to pass to type IIB. And then we apply S-duality. We arrive at the following final brane system shown in Figure 8.
\begin{center}
\begin{tabular}{c c c c c c}
Type IIB 10D ST  & $S^{1}$     & $\times$ & $T^{\ast}M^3 $ & $\times $  & $\real^3$\\
1 NS5 brane      & $\text{pt}$ & $\times$ & $T^{\ast}M^3 $ & $\times $  & $\lac 0 \rac$\\
M D3 branes      & $\text{pt}$ & $\times$ & $M^3 $         & $\times $  & $\real_{+}$\\  
N D3 branes      & $\text{pt}$ & $\times$ & $M^3$          & $\times $  & $\real_{-}$ 
\end{tabular}
\end{center}
The S-duality maps D5 brane to NS5 brane. The former was obtained from the above D6 brane. On the stack of the D3 branes, its worldvolume theory is $4d\hspace{0.1cm}\mathcal{N}=4$ super Yang-Mills with gauge groups $U(M)$ whereas the theory on the other brane stack has gauge group $U(N)$.
\newline

We next apply the (GL) topological twist along $M^3$ of $T^{\ast} M^3$ to the above super Yang-Mills theories~\cite{KW}. This results in a cohomological quantum field theory that is a coupled 4d-3d system across the NS5 brane. The cohomological sector of the theory is the Chern-Simons theory based on $U(N|M)$ (up to $Q$-exact terms). Its action functional is the supergroup Chern-Simons theory (up to certain exact terms). Furthermore, analogous to the Chern-Simons level parameter in case of a Lie group $SU(N)$, $U(N|M)$ Chern-Simons theory carries a parameter $K$, which comes from the complexified gauge coupling constant $\tau$ of the super Yang-Mills theory, which in turn comes from the complexified string coupling constant.
$$
\tau = K cos(\theta) e^{i \theta} \in H^{+},
$$
where $\theta$ is the vaccum angle and $H^{+}$ the upper half complex plane ($Im\, \tau > 0$). The $U(N|M)$ Chern-Simons theory is supported on $M^3$ in the NS5 brane.
Its action functional at level $K$ is
$$
CS(\mathcal{A}) = \frac{iK}{4\pi} \int_{M^3} Str \lb \mathcal{A}  d \mathcal{A} + \frac{2}{3} \mathcal{A}^3 \rb + \lac Q, \cdots \rac, 
$$
where $\mathcal{A} = \mathcal{A}_b + \mathcal{A}_f, \mathcal{A}_f$ is a fermion field and $\mathcal{A}_b$ is the complexified gauge connection of $A$ 
$$
\mathcal{A}_b = \lb A + i sin(\theta) \phi \rb |_{y=0^{\pm}},
$$
$+$ sign provides $G_r$-part of $\mathcal{A}_b$ whereas $-$ sign provides $G_l$-part of $\mathcal{A}_b$~\footnote{Recall that the bosonic (Grassman even) part of $U(N|M)$ is $ U(N) \oplus U(M)$.}.   
\newline

\noindent The existence of the super $\hat{Z}_{b,c}$ can be predicted from 11 dimensions. Specifically, the presence of the cigars in Figure 8, in particular their geometry away from the tips, requires imposing (asymptotic) boundary conditions $(b,c) \in H_1 (M^3 ;\intg)^N \times H_1 (M^3 ;\intg)^M$. The partition function over the BPS sector of the Hilbert space of the brane system is
$$
\hat{Z}^{gl(N|M)}_{b,c} [M^3 ; q] : = Tr_{H_{b,c}} (-1)^F q^{L_0}.
$$
where $F$ is fermion number operator and $L_0 $ is the generator of $U(1)_q$.

\section{Super series for knot complements}

Motivated by the idea of partial surgery (29), a generalization of (56) to complements of plumbed knots was analyzed in \cite{C5}. We found that a series invariant of plumbed knot complement associated with $sl(2|1)$ is a three variable series. Specifically, it is sum of contributions from good chambers $\alpha_{\pm}$.
$$
F_K (y,z,q) : = F_K (y,z,q;\alpha_+)  + F_K (y,z,q;\alpha_-),
$$
The general form of the super $F_K$ is
\begin{equation}
F_K  (y,z,q) = c + \sum_{ \substack{m,n \in \intg_{\geq 0}^2 \\ (m,n) \neq (0,0)}} f_{m,n}(K; q) \lb \frac{y^m}{z^n} - \frac{z^n}{y^m} \rb \quad \in \intg + q^{\Delta}\intg[q^{-1},q]] [[y/z, (y/z)^{-1} ]].
\end{equation}
It carries the Weyl symmetry $y \leftrightarrow y^{-1}$ and $z \leftrightarrow z^{-1}$. In comparison with (30), the summation of (30) is over odd integers whereas (61) is over a pair of nonzero integers.

\subsection{Torus knots}

We use the plumbing graph descriptions of torus knots in Section 3.1 to find good chambers for infinite families of the torus knots.
\begin{Prop} (\cite{C5}) Let $v$ be the number of vertices of plumbing graphs of $T(2, 2n+1)$ and $T(3, 3n+ w), w=1,2$ and $\alpha_{+}= (\alpha_1, \alpha_2 , \alpha_{v-1})$ and $\alpha_{-}$ be the good chambers for torus knots , where $\alpha_1$ corresponds to degree three vertex and the other two are associated with degree one vertices of their plumbing graphs. Their good chambers given by
$$
\alpha_{+} =  (1, 1, 1),\quad \alpha_{-} = -\alpha_{+},
$$
yield a well defined (Laurent) power series $f_{m,n}(q)$. 
\end{Prop}
\noindent The general structure of the super $F_K$ of torus knots $K=T(2,2l+1), l\geq 2$ splits into q independent and dependent parts. The former is a new feature in the super $F_K$, which is absent in (30).  And it can be expressed in terms of the unknot. The latter has the following form.
$$
\sum_{(m,n)\in \intg_{+}^2 } \epsilon_{m,n} q^{\frac{m(m+g(m,n))}{2(2l+1)}}\lb \frac{y^m}{z^{m+g(m,n)}} + \frac{y^{m+g(m,n)}}{z^{m}} -  \frac{z^m}{y^{m+g(m,n)}} - \frac{z^{m+g(m,n)}}{y^{m}}\rb,
$$
where $g(m,n;K)=g(m,n) \in \lac 1,3,5,\cdots, 2l-1 \rac$ and $\epsilon_{m,n} (K) = \epsilon_{m,n} $ is a sign function (see \cite{C5} for its algorithm).
\newline

\noindent\underline{Examples} We list a few examples of the torus knots (additional examples are recorded in \cite{C5}).
$$
F_{T(2,3)}(y,z,q)  =  1 + \sum_{i=2}^{\infty} \lb y^i +  \frac{1}{z^i} \rb - \sum_{i=2}^{\infty} \lb \frac{1}{y^i} + z^i \rb +q \left(\frac{y^2}{z^3}+\frac{y^3}{z^2}  - \frac{z^2}{y^3}-\frac{z^3}{y^2} \right)
$$
$$
+q^2 \left(\frac{y^3}{z^4}+\frac{y^4}{z^3} -\frac{z^3}{y^4}-\frac{z^4}{y^3}\right) +q^5 \left(-\frac{y^5}{z^6}-\frac{y^6}{z^5} +\frac{z^5}{y^6}+\frac{z^6}{y^5}\right)+q^7 \left(-\frac{y^6}{z^7}-\frac{y^7}{z^6} +\frac{z^6}{y^7}+\frac{z^7}{y^6}\right) 
$$
$$
 +q^{12}\left(\frac{y^8}{z^9}+\frac{y^9}{z^8} - \frac{z^8}{y^9}-\frac{z^9}{y^8}\right)  +q^{15} \left(\frac{y^9}{z^{10}}+\frac{y^{10}}{z^9} -\frac{z^9}{y^{10}}-\frac{z^{10}}{y^9}\right) + \cdots
$$

$$
F_{T(2,5)}(y,z,q) = 1 + \sum_{\substack{i=2 \\ i\neq 3}}^{\infty} \lb y^i + \frac{1}{z^i} \rb  - \sum_{\substack{i=2 \\ i\neq 3}}^{\infty} \lb \frac{1}{y^i} + z^i \rb+q \left(\frac{y^2}{z^5}+\frac{y^5}{z^2}-\frac{z^2}{y^5}-\frac{z^5}{y^2}\right)
$$
$$
+ q^2 \left(\frac{y^4}{z^5}+\frac{y^5}{z^4}-\frac{z^4}{y^5}-\frac{z^5}{y^4}\right)+q^3\left(\frac{y^5}{z^6}+\frac{y^6}{z^5}-\frac{z^5}{y^6}-\frac{z^6}{y^5}\right)+q^4 \left(\frac{y^5}{z^8}+\frac{y^8}{z^5}-\frac{z^5}{y^8}-\frac{z^8}{y^5}\right)
$$
$$
+q^7 \left(-\frac{y^7}{z^{10}}-\frac{y^{10}}{z^7} +\frac{z^7}{y^{10}}+\frac{z^{10}}{y^7}\right) +q^9 \left(-\frac{y^9}{z^{10}}-\frac{y^{10}}{z^9} +\frac{z^9}{y^{10}}+\frac{z^{10}}{y^9}\right) + \cdots
$$

$$
F_{T(3,4)}(y,z,q) = 1 + \sum_{\substack{i=3 \\ i\neq 5}}^{\infty} \lb y^i + \frac{1}{z^i} \rb  - \sum_{\substack{i=3 \\ i\neq 5}}^{\infty} \lb \frac{1}{y^i} + z^i \rb 
+q \left(\frac{y^3}{z^4}+\frac{y^4}{z^3}-\frac{z^3}{y^4}-\frac{z^4}{y^3}\right)
$$
$$
+q^2 \left(\frac{y^3}{z^8}+\frac{y^4}{z^6}+\frac{y^6}{z^4}+\frac{y^8}{z^3}-\frac{z^3}{y^8}-\frac{z^4}{y^6}-\frac{z^6}{y^4}-\frac{z^8}{y^3}\right)+q^3 \left(\frac{y^4}{z^9}+\frac{y^9}{z^4}-\frac{z^4}{y^9}-\frac{z^9}{y^4}\right)+q^4\left(\frac{y^6}{z^8}+\frac{y^8}{z^6}-\frac{z^6}{y^8}-\frac{z^8}{y^6}\right)
$$
$$
+q^6 \left(\frac{y^8}{z^9}+\frac{y^9}{z^8}-\frac{z^8}{y^9}-\frac{z^9}{y^8}\right)+q^7 \left(-\frac{y^7}{z^{12}}-\frac{y^{12}}{z^7}+\frac{z^7}{y^{12}}+\frac{z^{12}}{y^7}\right)+q^{10}\left(-\frac{y^{10}}{z^{12}}-\frac{y^{12}}{z^{10}}+\frac{z^{10}}{y^{12}}+\frac{z^{12}}{y^{10}}\right)
$$
$$
+q^{11} \left(-\frac{y^{11}}{z^{12}}-\frac{y^{12}}{z^{11}}+\frac{z^{11}}{y^{12}}+\frac{z^{12}}{y^{11}}\right)+q^{13}\left(-\frac{y^{12}}{z^{13}}-\frac{y^{13}}{z^{12}}+\frac{z^{12}}{y^{13}}+\frac{z^{13}}{y^{12}}\right)+q^{14}\left(-\frac{y^{12}}{z^{14}}-\frac{y^{14}}{z^{12}}+\frac{z^{12}}{y^{14}}+\frac{z^{14}}{y^{12}}\right)
$$
$$
+q^{17}\left(-\frac{y^{12}}{z^{17}}-\frac{y^{17}}{z^{12}}+\frac{z^{12}}{y^{17}}+\frac{z^{17}}{y^{12}}\right) + \cdots
$$

\subsection{The Dehn surgery formula}

We provide the Dehn surgery formula relating the super $F_K$ to the super $\hat{Z}_{b,c}$.
\begin{Thm} (\cite{C5}) Let $Y_K$ be the complement of a knot $K$ in the 3-sphere $S^3$ and let $Y_{p/r}$ be a result of Dehn surgery along $K$ with slope $p/r \in \rational^{\ast}$. Assume that $Y_K$ and $Y_{p/r}$ are represented by negative definite plumbings. Then the invariants of $Y_{p/r}$ are given by
$$
\hat{Z}_{b,c}[Y_{p/r};q] = (-1)^{\tau} \mathcal{L}^{(\alpha_i ;\, p/r )}_{b,c} \lsb F^{(\alpha_i)}_K (y,z,q) \rsb,
$$
where the Laplace transform for $\alpha_+ $ chamber is
$$
\mathcal{L}^{(\alpha_+ ;\, p/r )}_{b,c} : y^{\alpha}z^{\beta}q^{\gamma} \mapsto q^{\gamma}
\begin{cases}
\sum\limits_{r_s = r_{s,min}}^{\infty}  q^{\frac{\beta(r \alpha + \epsilon r_s )}{p}}, & \text{if}\quad r \alpha + \epsilon r_s + b \in p\intg,\, r\beta + c \in p\intg\\
\sum\limits_{w_s = w_{s,min} }^{\infty} q^{\frac{\alpha(r \beta - \epsilon w_s )}{p}}, & \text{if}\quad   r \beta - \epsilon w_s + c \in p\intg ,\, r\alpha + b \in p\intg\\
0, & \text{otherwise}
\end{cases}
$$
and the Laplace transform for $\alpha_- $ chamber is 
$$
\mathcal{L}^{(\alpha_- ;\, p/r )}_{b,c} : y^{\alpha}z^{\beta}q^{\gamma} \mapsto -q^{\gamma}
\begin{cases}
\sum\limits_{w^{\prime}_s =  w^{\prime}_{s,min}  }^{\infty} q^{\frac{\beta(r \alpha - \epsilon w^{\prime}_s )}{p}}, & \text{if}\quad   r \alpha - \epsilon w^{\prime}_s + b \in p\intg ,\, r\beta + c \in p\intg\\
\sum\limits_{r^{\prime}_s =  r^{\prime}_{s,min} }^{\infty}  q^{\frac{\alpha(r \beta + \epsilon r^{\prime}_s )}{p}}, & \text{if}\quad r \beta + \epsilon r^{\prime}_s + c \in p\intg,\, r\alpha + b \in p\intg\\
0, & \text{otherwise}
\end{cases}
$$
where $r_{s,min},\, r^{\prime}_{s,min} \geq 1, \, w_{s,min},\, w^{\prime}_{s,min} \geq 0$ and $\epsilon = sign(p) (-1)^{\pi + 1}$.
\end{Thm}
We observe a qualitative difference between the above surgery formula and the $sl(2)$ surgery formula (40) in Section 3.5. For applications of Theorem 2.57, $S^3_{-1/r}(T(s,t))$ and $S^3_{-p}(T(s,t))$ for some values of $s,t,r,p$ and $S^3_{-p}(unknot)$ were considered in \cite{C5}.
\newline

\noindent\textbf{Future directions} We list open problems.

\begin{itemize}
	\item Obtaining an analytic formula of $\hat{Z}_b (q)$ defined on $|q|>1$ for positive definite plumbed manifolds $-Y_{\Gamma}$ has been a major challenge. Several approaches for computing $\hat{Z}_b (-Y_{\Gamma}, q)$ was listed in Section 2.7. However, they are applicable to specific examples of $-Y_{\Gamma}$.
	
	\item There are multiple Dehn surgery formulas (cf. Section 3.5). Each has restricted applicability. We believe that an unifying surgery formula exists and would be important.
	
	\item Extending the definition of $F_L$ beyond the homogeneous links would be valuable for conceptually and computationally.
		
\item A construction of the $\mathcal{H}_{BPS}$ in Section 2.2 for categorification of the WRT invariant is highly desirable.
	
\end{itemize}

\end{document}